# On the effects of transformation strain induced by hydride precipitation

Masoud Taherijam[1], Saiedeh Marashi[1], Alireza Tondro[1,2], Hamidreza Abdolvand[1*]

[1] *Department of Mechanical & Materials Engineering, Western University, London, Ontario, Canada, N6A 5B9*

[2]*Materials and Fuel Performance Testing, Canadian Nuclear Laboratories, Chalk River, Ontario, Canada, K0J 1J0*

---

[*] Corresponding author: H. Abdolvand (hamid.abdolvand@uwo.ca)

# On the effects of transformation strain induced by hydride precipitation

Masoud Taherijam[1], Saiedeh Marashi[1], Alireza Tondro[1,2], Hamidreza Abdolvand[2*]

[1] *Department of Mechanical & Materials Engineering, Western University, London, Ontario, Canada, N6A 5B9*

[2] *Materials and Fuel Performance Testing, Canadian Nuclear Laboratories, Chalk River, Ontario, Canada, K0J 1J0*

**Abstract**
One of the main degradation mechanisms of the zirconium alloys used in nuclear reactors is hydrogen embrittlement and hydride formation. The formation of zirconium hydrides is accompanied by a transformation strain, the effects of which on the development of localized deformation zones are not well-understood. This study uses a crystal plasticity finite element model that is coupled with diffusion subroutines to quantify such effects. For this purpose, a zirconium specimen was hydrided in the absence of any external mechanical loads. With the use of electron backscatter diffraction, the rotation fields around interacting intragranular hydrides as well as those located at grain boundaries or triple points were measured at a high spatial resolution. The as-measured zirconium and hydride morphologies were mapped to the model for numerical simulation. Both numerical and experimental results show that hydride precipitation induces large rotation fields within the zirconium matrix, where such rotations are at their maximum in the vicinity of hydride tips. While the crystallographic orientations and shapes of hydrides affect the magnitude of rotation fields, both experimental and modeling results revealed the development of discrete and parallel geometrically necessary dislocation fields and a strong interaction among neighboring hydrides. It is shown that the stress field resulting from hydride precipitation affects the patterning of hydrogen distribution, which in return affects further hydride interlinking.

* Corresponding author: H. Abdolvand (hamid.abdolvand@uwo.ca)

## 1. Introduction

Hydrogen embrittlement and hydride formation are among the main degradation mechanisms of zirconium alloys used in thermal nuclear reactors. These alloys tend to corrode and absorb hydrogen when exposed to water, with hydrogen atoms subsequently diffusing into the zirconium matrix. During maintenance, the temperature of zirconium-alloy components decreases, reducing the hydrogen solubility limit which leads to the precipitation of hydrides. This can result in a reduction of fracture toughness, and possible failure of critical components [1,2]. One of the key failure modes associated with hydride formation is Delayed Hydride Cracking (DHC), which occurs when hydrides become sufficiently large and crack within the surrounding zirconium matrix [3,4]. Therefore, understanding the mechanisms behind hydrogen embrittlement and hydride formation is crucial for ensuring the safety of nuclear reactors.

The formation of hydrides is a complex phenomenon that is affected by several factors, including the microstructure, e.g., grain size and texture, the thermal history, the applied and state of residual stresses, as well as the hydrogen content of zirconium matrix [5–12]. There have been several numerical and experimental studies investigating the precipitation of hydrides in zirconium alloys, for which comprehensive reviews are conducted and can be found in [13,14]. The process of hydride formation involves a transformation strain that arises due to the lattice mismatch between the hydride precipitate and the zirconium matrix. This strain can lead to the anisotropic elastic and plastic deformation of the material [15–17]. As a result, hydrides can induce large rotations in the surrounding zirconium matrix, leading to the development of localized deformation zones. These deformation zones can promote the formation of cracks and ultimately lead to material failure. Theoretical stress-free transformation strains were calculated by Carpenter [18], while Barrow et al. [19] experimentally measured the interfacial strains of precipitated hydrides using nano-beam electron diffraction. Although there is a discrepancy between the experimental and theoretical results, the experimental values were obtained from hydrides that had undergone plastic relaxation, as suggested by Blackmur et al. [20]. This highlights the significance of considering the effect of plastic deformation on the observed transformation strains in hydrides.

Various aspects of hydride formation and growth are studied through the use of electron microscopy [21–23]. For example, transmission electron microscopy observations show that hydrides can form at the nanoscale, and their morphologies depend on the crystallographic orientation of the surrounding zirconium matrix [24,25]. The delta (δ) phase is the most frequently observed hydride in zirconium alloys [26], which has a Face-Centered Cubic (FCC) crystal structure. δ-hydrides have a well-defined orientation relationship with the Hexagonal Close-Packed (HCP) α-zirconium matrix, i.e., $\{111\}_\delta \parallel \{0002\}_\alpha$ and $<110>_\delta \parallel <11\bar{2}0>_\alpha$ [27]. This alignment between the crystallographic planes and directions of the two phases is crucial for the nucleation and growth of δ-hydrides. The hydrides can form either as inter-granular or intra-granular precipitates depending on the cooling rate [28]. Electron BackScatter Diffraction (EBSD) has also been widely used to investigate microstructural variations of α-grains [29,30], or microstructural effects during hydride formation. For instance, Wang et al. [31] used EBSD to study the orientation dependence of hydride precipitates in commercially pure titanium. Long et al. [21] used EBSD to study the growth of hydride blisters in a Zircaloy-4 plate. Mani Krishna et al. [32] used EBSD technique to characterize the role of α/β interface on hydride formation in Zr-2.5 Nb. Their analysis showed that hydrides have higher preference to form along α/β interfaces. Marashi et al. [30] used high spatial resolution EBSD and imaging to resolve local microstructure variations in Zr-2.5Nb

CANDU pressure tubes. Due to the presence of a crystallographic relationship between hydrides and parent α-grains, such variations affect the patterning of hydrides. While with the developed procedure it was possible to measure the orientations of β-crystals, it was shown that microstructure variations in Zr-2.5Nb are non-negligible. Kiran Kumar et al. [33] utilized EBSD to investigate the morphology and orientation of hydrides within a cold-worked and stress relieved zircaloy-4 specimen.

While early studies assumed that hydrides are brittle [34,35], further investigations have provided some insight into the complex process of slip transmission from the zirconium matrix to the hydride phase [36–38]. For example, in a study conducted by Weekes et al. [39] on Zircaloy-4 micropillars containing hydrides, it was shown that zirconium hydrides might deform plastically under a uniaxial load. It was further shown that shear bands were initiated in the zirconium matrix before propagating into the hydride domain. Wang et al. [40] used in-situ high resolution EBSD to analyze the evolution of stress and Geometrically Necessary Dislocations (GND) densities in Zircaloy-4 microcantilevers, both with and without hydrides. An increased GND density in hydrided samples was observed. In another study, Wang et al. [41] used in situ SEM micropillar compression tests to show that certain orientations favor shear at the hydride–matrix boundary, while others require increased shear stress to initiate slip in hydrides. Similar to zirconium, hydride formation is also observed in titanium alloys. It has been shown that thin titanium hydrides can sustain cyclic plastic strains and accommodate up to 70% shear deformation [42–46]. Further, it was shown by Conforto et al. [47] that titanium hydrides exhibit ductility when they are thinner than 400 nm. The factors governing the transition of hydrides from ductile to brittle, however, are yet to be fully understood.

In recent years, different numerical techniques have been used to model the behavior of zirconium alloys in the presence of hydrogen and hydrides. Several studies investigated the behavior of hydrogen and the formation of hydrides in zirconium using density functional theory [48–50]. Szewc et al. [51] studied the effects of hydride formation on the onset of plasticity in zirconium using molecular dynamics simulations. The simulations revealed that plasticity is always initiated by the nucleation and propagation of pyramidal partial dislocations from surfaces, followed by the formation of basal dislocations. Patel et al. [52] devised a multiscale method to model the precipitation and reorientation of hydrides in a state of equilibrium. Some researchers employed discrete dislocation dynamics modeling to investigate the stress fields around hydrides in zirconium [53–57]. While this approach is capable of considering the interactions between individual dislocations, it is computationally expensive and does not account for the time scale of the phenomenon being studied. In contrast, Crystal Plasticity Finite Element (CPFE) is a mesoscale modeling method that can capture the real-time deformation of individual grains or clusters of grains by incorporating the effects of plastic slip on active slip systems. For example, CPFE modeling has been used to investigate the effects of crystallographic texture, grain size, and loading conditions on the plastic deformation of zirconium alloys [58–64]. Liu et al. [65] investigated hydride formation in Zircaloy-4 under cyclic thermomechanical loading using microstructurally representative CPFE modeling. CPFE can also be coupled with diffusion modules to understand the stress-assisted diffusion of hydrogen atoms [66]. For example, Ilin [67] used a coupled Diffusion-CPFE (D-CPFE) approach to study the diffusion of hydrogen atoms within stainless steels and highlighted the crucial effects of hydrostatic stress and strain rate on the redistribution of hydrogen atoms. Furthermore, the contributions of texture and grain morphology to hydrogen transport towards stress-risers, such as notches, were deconvoluted in the study by Tondro and Abdolvand [68]. The effects of GNDs on trapping of hydrogen atoms and their contribution to the total concentration of hydrogen atoms were also studied by Tondro et al. [69] using a CPFE approach. GNDs are a result of the incompatibility in crystal lattice due to the presence of impurities or defects in the material.

While several studies have focused on the effect of hydride precipitation on the local plasticity induced in the zirconium matrix, less attention is given to the measuring or modeling the corresponding rotation field, particularly within the hydrides. Hence, the focus of this study is to characterize the deformation field induced by the precipitation of hydrides in the absence of external mechanical loads. Both rotation and GND fields within the zirconium matrix as well as hydrides were characterized using high spatial resolution EBSD. The "as measured" grain morphologies and orientations were mapped into a CPFE model to simulate the distribution of GND densities and misorientation fields and compare them with those from EBSD measurements. The combination of experimental measurements and modeling allowed conducting a comprehensive analysis of the effects of hydrides transformation strain on the patterning of local stress, rotation, and hydrogen concentration fields.

## 2. Methodology
### 2.1 Sample preparation and measurement set-up

A pure zirconium bar was used which comprised of 2500 ppm Hf, 1000 ppm O, 250 ppm C, 200 ppm Cr, 200 ppm Fe, 100 ppm N, and 10 ppm H. This bar was firstly heat-treated at 730 °C for 48 hours to remove prior residual stresses and recrystallize large grains. The average grain size of the heat-treated bar was more than 0.7 mm (see Fig. 1). A thick specimen was cut from the heat-treated bar, polished, and then cathodically charged in a 0.2 mol solution of $H_2SO_4$ at 65 °C for 48 hours. The charged specimen was subsequently homogenized at 450 °C for 2 hours in an Argon environment, and slowly cooled using 1K/min to room temperature to have δ-hydrides forming both within α-grains and at grain boundaries. The hydrided specimen was mechanically polished down to 1200 grit, and then in a dilute solution of 1 µm, 0.5 µm, and 0.2 µm of Alumina. To acquire high quality diffraction patterns, the specimen was finally electropolished for 60 sec at 20 Vol in a -35 °C solution of 10% perchloric acid and 90% methanol. A low-magnification image of the electropolished specimen is shown in Fig. 1. Hydrides are observed to precipitate both at grain boundaries and within α-grains.

EBSD measurements were carried out using an Apreo-2 Field Emission Gun Scanning Electron Microscope (FEG-SEM), equipped with a Bruker high-resolution EBSD system. For performing EBSD scans, a 20 keV electron beam with a 6.5 nA probe current was used at the working distance and sample-to-detector distance of 18 mm. All measurements were done using the pattern resolution of 400x300 pixels, and the same step size of 40 nm to ensure consistency in the calculation of GNDs. The MTEX software [70] was used for the analysis of Euler angles and for the determination of GND densities (see below). The EBSD scans were conducted on three different types of hydrides including intragranular, grain boundary, and triple point hydrides (see Fig. 1).

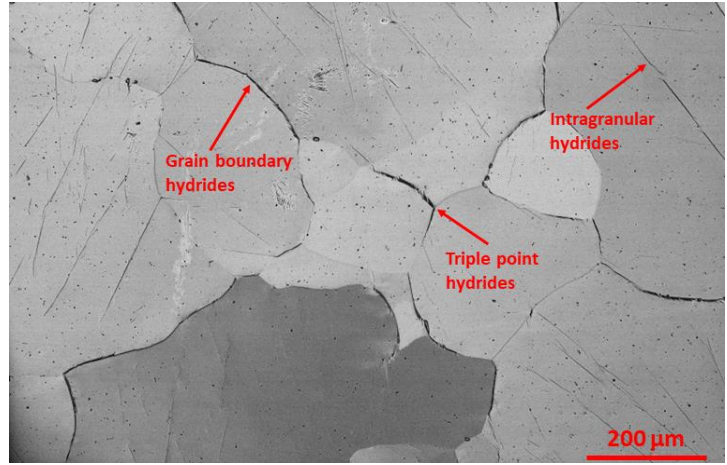

**Fig. 1. A low magnification image of the hydrided specimen with large α-grains.**

## 2.2 CPFE model formulation

The numerical simulations of this study were conducted using a coupled Diffusion-CPFE model (D-CPFE). This section outlines the key equations used in the model, but further details can be found elsewhere [66]. A User MATerial (UMAT) subroutine is coupled with a UMATHT subroutine for simulating the diffusion of hydrogen atoms in an anisotropic polycrystal. The UMAT subroutine calculates stress increments and updates the solution dependent state variables based on the inputs provided by the FE solver, i.e., strain, rotation, time, and temperature increments. The total strain increment ($\Delta \boldsymbol{\varepsilon}$) can be written as:

$$\Delta \boldsymbol{\varepsilon} = \Delta \boldsymbol{\varepsilon}^{el} + \Delta \boldsymbol{\varepsilon}^{pl} + \Delta \boldsymbol{\varepsilon}^{trH} \tag{1}$$

where $\Delta \boldsymbol{\varepsilon}^{el}$ and $\Delta \boldsymbol{\varepsilon}^{pl}$ respectively represent the increments of elastic strain and the plastic strain resulting from crystallographic slip. $\boldsymbol{\varepsilon}^{trH}$ is the transformation strain from the precipitation of hydrides, which is incrementally applied to the hydride domain. For calculating the plastic strain increment from slip, the plastic part of deformation rate ($\boldsymbol{D}^{pl}$) tensor is integrated over each time increment, i.e.:

$$\dot{\boldsymbol{\varepsilon}}^{pl} = \boldsymbol{D}^{pl} = \sum_{\alpha=1}^{N^{spl}} \boldsymbol{P}^{\alpha} \dot{\gamma}^{\alpha} \tag{2}$$

$$\boldsymbol{P}^{\alpha} = sym(\boldsymbol{S}^{\alpha}) \text{ where } \boldsymbol{S}^{\alpha} = \vec{d}^{\alpha} \otimes \vec{n}^{\alpha}$$

where $\boldsymbol{P}^{\alpha}$ is the symmetric part of the Schmid tensor ($\boldsymbol{S}^{\alpha}$), $\dot{\gamma}^{\alpha}$ is the shear strain rate on the slip system $\alpha$, and $\vec{d}^{\alpha}$ and $\vec{n}^{\alpha}$ are respectively slip direction and normal to the slip plane for the same slip system. The following equation is used to calculate shear strain rates [71]:

$$\dot{\gamma}^{\alpha} = \dot{\gamma}_0 \left| \frac{\tau^{\alpha}}{g^{\alpha}} \right|^n sgn\left(\frac{\tau^{\alpha}}{g^{\alpha}}\right) \tag{3}$$

where $\dot{\gamma}_0$ is a reference shear strain rate and $n$ controls the rate dependency. $\tau^{\alpha}$ and $g^{\alpha}$ are the resolved shear stress and the strength of slip system $\alpha$, respectively, and $sgn(.)$ denotes the sign function. An extended Voce hardening law is used to calculate the strength of slip systems:

$$g^\alpha = g_0^\alpha + (g_1^\alpha + \theta_1^\alpha \Gamma)\left(1 - exp\left(-\frac{\theta_0^\alpha \Gamma}{g_1^\alpha}\right)\right) \tag{4}$$

where $g^\alpha$ is the current Critical Resolved Shear Stress (CRSS), $g_0^\alpha$ is the initial CRSS, $\Gamma$ is the shear accumulated on all slip systems, $\theta_0^\alpha$ is the initial hardening rate, and $g_1^\alpha$ and $\theta_1^\alpha$ determine asymptotic characteristics of hardening.

The resolved shear stress used in Eq. 3 is proportional to the Kirchhoff stress ($\boldsymbol{\Psi}$) through [72]:

$$\tau^\alpha = \boldsymbol{P}^\alpha : \boldsymbol{\Psi} \tag{5}$$

The Jaumann rate of Kirchhoff stress ($\boldsymbol{\breve{\Psi}}$) correlates with the elastic part of deformation rate tensor ($\boldsymbol{D}^{el}$), and the elastic stiffness tensor ($\mathbb{C}$) through the following equation:

$$\boldsymbol{\breve{\Psi}} = \mathbb{C} : \boldsymbol{D}^{el} \text{ where } \boldsymbol{\breve{\Psi}} = \boldsymbol{\dot{\Psi}} - \boldsymbol{\Omega}^{el}\boldsymbol{\Psi} + \boldsymbol{\Psi}\boldsymbol{\Omega}^{el} \tag{6}$$

where $\boldsymbol{\Omega}^{el}$ is the elastic part of spin tensor, which can be calculated by subtracting the plastic part of spin tensor ($\boldsymbol{\Omega}^{pl}$) from the total spin tensor. The plastic part of spin tensor can be determined using the following equation:

$$\boldsymbol{\Omega}^{pl} = \sum_{\alpha=1}^{N^{spl}} \boldsymbol{W}^\alpha \dot{\gamma}^\alpha \text{ where } \boldsymbol{W}^\alpha = asym(\boldsymbol{S}^\alpha) \tag{7}$$

The unconstrained misfit strain associated with the formation of δ-hydrides were calculated to be 4.58% and 7.2% along the HCP crystal a-axis and c-axis, respectively [18]:

$$\boldsymbol{\varepsilon}^{trH} = \begin{bmatrix} 0.0458 & 0 & 0 \\ 0 & 0.0458 & 0 \\ 0 & 0 & 0.072 \end{bmatrix} \tag{8}$$

This transformation strain was applied incrementally and slowly to mimic the quasi-static condition of the domain that undergoes precipitation.

### 2.2.1 Stress-assisted hydrogen diffusion

The flux of hydrogen atoms ($\vec{J}$) within zirconium matrix depends on the concentration of hydrogen atoms (C), temperature and stress gradients:

$$\vec{J} = -\boldsymbol{D} \cdot \left[\vec{\nabla}C + C. lnC \frac{\vec{\nabla}T}{T} - C\frac{\bar{V}_H}{3.RT}\vec{\nabla}\sigma_{kk}\right] \tag{9}$$

where $\boldsymbol{D}$ is the diffusivity of hydrogen atoms in the zirconium matrix, $\bar{V}_H$ is the partial molar volume of hydrogen atoms in the Zr lattice and is 1670 mm³/mol, $\sigma_{kk}$ is hydrostatic stress times three, and R is the gas constant. The three terms in the right-hand side of equation (9) respectively represent hydrogen diffusion due to concentration gradients ($\vec{\nabla}C$), temperature gradients ($\vec{\nabla}T$), and stress gradients ($\vec{\nabla}\sigma_{kk}$). Considering the size of the investigated area with EBSD and considering that a very slow cooling rate was used, the effects of temperature gradients can be disregarded from Eq. 9, i.e., we only focus on the effects of stress on mass diffusion and solve for an isothermal case where the gradient of temperature is zero.

The diffusivity coefficients of hydrogen atoms along the a-axis ($D_a$) and c-axis ($D_c$) of the HCP zirconium crystals for the temperature range of 300K-1100 K are [73]:

$$D_c = 1.08 \exp\left(-\frac{0.46}{8.6 \times 10^{-5}T}\right)$$
$$D_a = D_c(-3.3 \times 10^{-7}T^2 + 7 \times 10^{-4}T + 0.8298)^{-1}$$
(10)

in which diffusivities are in units of mm/s$^2$. The general form of diffusion equation derived from a control volume can be written as:

$$\frac{\partial C}{\partial t} = -\nabla \cdot \vec{j}$$
(11)

For the steady state condition, the derivative of concentration with respect to time will be zero.

The format of Eqs. 9-11 is very similar to the conductive heat transfer equations used for calculating temperature distribution in the presence of heat sinks and sources. In Abaqus FE solver, UMATHT can be used for modeling coupled temperature-displacement problems, but with replacing temperature with concentration (C), this subroutine can be used for coupled diffusion-displacement problems. Following the method discussed in [66], Eqs. 9-13 are implemented in a UMATHT subroutine to solve for hydrogen concentration.

To summarize, the total strain and time increments provided at the beginning of each time increment are used to calculate the stress increment using Eqs. 1-7. If a domain undergoes hydride precipitation, the transformation strain provided in Eq. 8 is applied to the pre-defined hydride domain, after applying appropriate rotations extracted from EBSD. The updated stress matrix at each IP is then used to calculate hydrostatic stress, and with the use of current coordinates of the same IP, the gradients of hydrostatic stress are calculated. The calculated $\vec{\nabla}\sigma_{kk}$ is subsequently sent to UMATHT to calculate hydrogen concentration using Eqs. 9-11.

### 2.2.2 CPFE input models

The as-measured EBSD maps are imported into the CPFE model to simulate the effects of transformation strain and confirm the source of observed trends in the experiment. First, the measured morphologies of α-zirconium and hydrides were imported into the model and a biased mesh was generated to capture the details of hydrides shapes. Due to the 3D nature of the problem, the model was then extruded by 10 µm in the thickness direction to simulate a quasi-3D microstructure. Second order quadratic brick elements (C3D20R) were used to mesh the model, with a significantly higher mesh density applied in the region surrounding the hydrides (see Fig. 2). The measured Euler angles from EBSD were assigned to each mapped grain. Since hydrides were formed in the absence of any external mechanical loads, no external load was applied to the model, but the left and bottom surfaces were fixed from the movements in the $x$ and $y$ directions ($U_x = U_y = 0$), respectively (see Fig. 2). In our experiment, the specimen was polished after hydriding. That is, the analyzed hydrides would have formed under the ~1 mm surface that was removed during the polishing steps. Hence, it was assumed that hydrides were formed under the plane strain condition by restricting any movements on both front and back surfaces of the model ($U_z = 0$), yet this boundary condition was removed in the last step of the simulation to replicate the release of stresses caused by polishing. It is worth mentioning that further analysis using other types of boundary conditions, including the plane stress condition, indicated that the trends observed for misorientation fields do not vary when switching between the plane strain to the plane stress condition, yet a slightly higher rotations were calculated in the absence of the relaxation step for the case of plane strain boundary condition. Overall, CPFE simulations consistently demonstrated the same trends, regardless of the specific boundary conditions applied. Additional analysis concerning the effects of boundary conditions can be found in the supplementary file. Since α-zirconium parents are very large grains, buffer layers were added to the model to minimize the effects of boundary conditions on the deformation fields that develop due to the hydride transformation strain. In all models, the buffer layer is much bigger than hydrides in both $x$ and $y$ directions, have the same material properties as the α parent grain, and the numerical results confirmed the absence of interaction between the hydrides and boundaries.

Following the method proposed by Tondro et al [74], the hydride transformation strain was applied to the elements of hydride domains incrementally, over 3600 seconds to replicate a quasi-static loading condition. If more than one hydride was modeled within a map, the transformation strains for all hydrides were applied simultaneously, in one simulation step. Further analysis showed that the trends reported in this paper are not sensitive to the sequence of hydriding, but the magnitudes may be slightly affected. The single crystal hardening parameters used for α-zirconium were the ones reported by Abdolvand et al [62] and provided in Table 1. The single crystal elastic constants that were used for α-zirconium were the ones determined by Fisher and Renken [75]: $C_{11}$=143.5 GPa, $C_{33}$=164.9 GPa, $C_{12}$=72.5 GPa, $C_{13}$=65.4 GPa, and $C_{44}$=32.1 GPa. For hydrides, the elastic constants of FCC δ-hydrides reported by Olsson et al. [76] were used: $C_{11}$=162 GPa, $C_{12}$=103 GPa, $C_{44}$=69.3 GPa. While determining the plastic properties of hydrides is not in the scope of current study, many different combinations of hardening parameters were used starting from relatively soft to a fully elastic medium. The results of these studies are presented in discussion section 4.1. For the sake of consistency, in all results presented in section 3, plastic deformation on {111}<110> slip system was allowed in the hydride domain using the CRSS value of 340 MPa which evolves linearly with the total plastic shear accumulated on all slip systems (Γ) with the slope of 90 MPa. In addition, although hydride precipitation takes place in a temperature window of about 150 ˚C, all results presented in this paper are based on the CRSS values presented in Table-1. Further CPFE simulations using temperature-dependent elastic constants and CRSS values revealed minor effects of temperature variations on the results presented.

Table 1- The single crystal parameters used for α-Zr

| Slip system | n | $\dot{\gamma}_0$ (s$^{-1}$) | $g_0^\alpha$ (MPa) | $g_1^\alpha$ (MPa) | $\theta_0^\alpha$ (MPa) | $\theta_1^\alpha$ (MPa) |
|---|---|---|---|---|---|---|
| Prism | 20 | $3.5 \times 10^{-4}$ | 120 | 330 | 10 | 0 |
| Basal | 20 | $3.5 \times 10^{-4}$ | 168 | 220 | 50 | 0 |
| Pyramidal <c+a> | 20 | $1.0 \times 10^{-4}$ | 330 | 270 | 620 | 280 |

For calculating hydrogen concertation, diffusion subroutines were called after applying transformation strain. Since the cooling rate was 1K/min and considering the diffusion length of hydrogen, all results presented for hydrogen concentrations are for the steady state condition. Since it was not feasible to measure local hydrogen concentrations, no comparison is made between CPFE and experimental results, yet hydrogen concentrations are solely provided to discuss the effects of localized stresses on the concentration of hydrogen atoms. A boundary condition of 100 wt.ppm was applied to the side surfaces of the model to study the depletion and accumulation of hydrogen atoms around hydrides. This boundary condition was applied because the solid solution solubility limit of hydrogen in the zirconium lattice at 300 ˚C, the nominal operating temperatures of nuclear reactors, is about 100 wt.ppm [13,15,77]. Although some DFT calculations conducted at nanoscales suggest that concentrations below 300 ppm do not provide enough driving force for the formation of γ and δ hydrides [78]. In addition, atom probe tomography (APT) results of [79] has revealed an unexpectedly higher deuterium atomic ratio of approximately 10%, surpassing the maximum solubility limit of about 2%. This difference might be attributed to the distinct maximal solubility under specific electrochemical-charging conditions or to the influence of defects near the polished sample surface [80]. APT analysis has also confirmed the presence of a layer of ζ-hydride that encapsulate δ-hydrides. This supports the core-shell morphology and points to a lower stoichiometric number for the transition from α-zirconium to δ-hydrides [81,82]. It is crucial to note that these effects occur at length scales much smaller than the one used in this paper. In addition, our primary goal is to understand the redistribution of hydrogen atoms within zirconium lattice and in the vicinity of hydride domains. By employing the D-CPFE model, we gain valuable insights into patterning of hydrogen atoms in the presence of the stress fields induced by hydride precipitation.

To summarize the following steps were followed during numerical modeling:

- Step-1: The hydride transformation strain was applied. For this step, displacements along the x and y of the left and bottom surfaces were respectively set to zero, so as the displacement along the z direction of both front and back surfaces. The diffusion subroutines were called at the end of this step.
- Step-2: The displacement along the z direction of the front surface was removed to allow relaxation.

In both numerical and experimental results, Grain Reference Orientation Deviation (GROD) is calculated based on the misorientation between each point with respect to the mean orientation calculated for all points within the same grain. To provide a like-to-like comparison and to be consistent, all results presented for GRODs and GNDs are for Step-2 of modeling, after relaxation. Although, stress components are provided at the end of Step-1 to highlight their condition before "polishing" the specimen. In addition, the $x$, $y$, and $z$ of the coordinate system used respectively point to the east, south, and perpendicular to the plane, unless otherwise stated. In all modeling results, the buffer layer is removed to only focus on the local variations around hydrides.

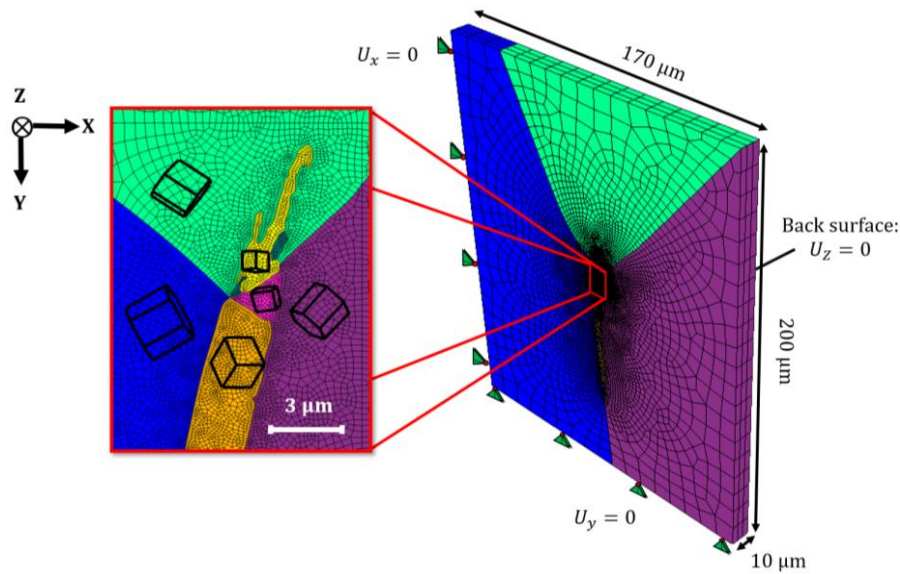

**Fig. 2. An example of CPFE input models for triple point hydrides built based on the measured grain orientations and morphologies of Fig. 12.**

### 2.2.3 The determination of geometrically necessary dislocation densities

Several methods can be used for the determination of GND densities within CPFE framework. A comprehensive comparison between the results of each method and those from EBSD is provided in [83]. These CPFE-based methods benefit from using the gradients of plastic shear on each slip system for the determination of GNDs- a parameter that is not available in an EBSD measurement. Hence, to provide a like-to-like comparison, the method proposed by Pantleon [84] was used for both CPFE and EBSD orientation maps. That is, the calculated Euler angles at the end of numerical simulations as well as the position of integration points were fed to the same subroutine that the measured EBSD maps were fed to. This subroutine is a part of the MTEX [70] toolbox for orientation analysis and uses the following procedure for determining GND densities.

The GND density $\rho^t$ can be calculated as [85]:

$$\alpha_{ij} = \sum_{t=1}^{N} b_i^t l_j^t \rho^t \quad \text{where } i,j = 1,2,3 \tag{12}$$

where $b_i^t$ and $l_j^t$ are the components of Burger's vector and dislocation line vector, respectively, N is the total number of possible dislocation types, and $\alpha_{ij}$ represents the components of Nye's tensor. The Nye's tensor can be written as a function of lattice curvature [85]:

$$\alpha_{ij} = \kappa_{ji} - \delta_{ji}\kappa_{mm} \tag{13}$$

In the above equation $\kappa_{ji}$ represents the components of the curvature tensor, with $\kappa_{ji} = \frac{\partial \theta_j}{\partial x_i}$, where $\theta_j$ is the lattice rotation vector and $\delta_{ji}$ is the Kronecker delta. In this paper, results for the total GND density are presented, i.e., $\rho_{GND} = \sum_{t=1}^{N} \rho^t$. Further details of GND calculation are available in the work by Pantleon [84].

### 3. Results

In this section, the interactions of intragranular hydrides are studied for two individual hydrides and a pack of hydrides. This is followed by an investigation into grain boundary and triple point hydrides.

### 3.1 Interacting intragranular hydrides

Two intragranular δ-hydrides embedded within an α-grain are shown in Fig. 3. The measured orientations of the α-grain and δ-hydrides are shown schematically in Fig. 3a with further orientation analysis provided in Fig. 3b. The c-axis of the α parent grain is oriented towards the x-axis, while two different orientations are measured for the two hydrides. It is evident from the orientation analysis that, the {111} planes and $< 110 >$ directions of both hydrides respectively coincide with the basal plane and the $< 11\bar{2}0 >$ direction of the α parent grain, confirming the orientation relationship reported for δ-hydrides. Fig. 3c shows the band contrast map for the as-measured EBSD map.

The EBSD measured and CPFE calculated GRODs are respectively shown in Fig. 3d and 3e. A very good agreement between the modeling and experimental results is achieved. It is evident that the transformation strain from hydrides induces large lattice rotations, as much as 2°, in the area in between the two hydrides. Interestingly, the extent and shape of rotation fields from model and experiment are identical, where the rotations are only observed in between the two hydrides. The measured and calculated GND maps are respectively shown in Fig. 3f and 3g. The results for the hydride domains are excluded from analysis due to being very thin (< 400 nm) and that more measurements points were required for quantifying variations. The variations within thicker hydrides are provided and discussed in section 3.3. It can be seen from the maps that GNDs are concentrated in the vicinities of hydride boundaries, both along the hydride and at the hydride tips. Interestingly, two highly concentrated parallel GND zones are identified in both CPFE and EBSD results which are highlighted with the black ovals of Fig. 3f. The GNDs are at their maximum, ~$10^{15}$ m$^{-2}$, along two parallel lines in between the hydrides, further highlighting the degree of interaction. The two parallel lines in the GND maps coincide with the prism slip system of the parent grain (see Fig. 3a), and are located at the places where slip activity is the highest.

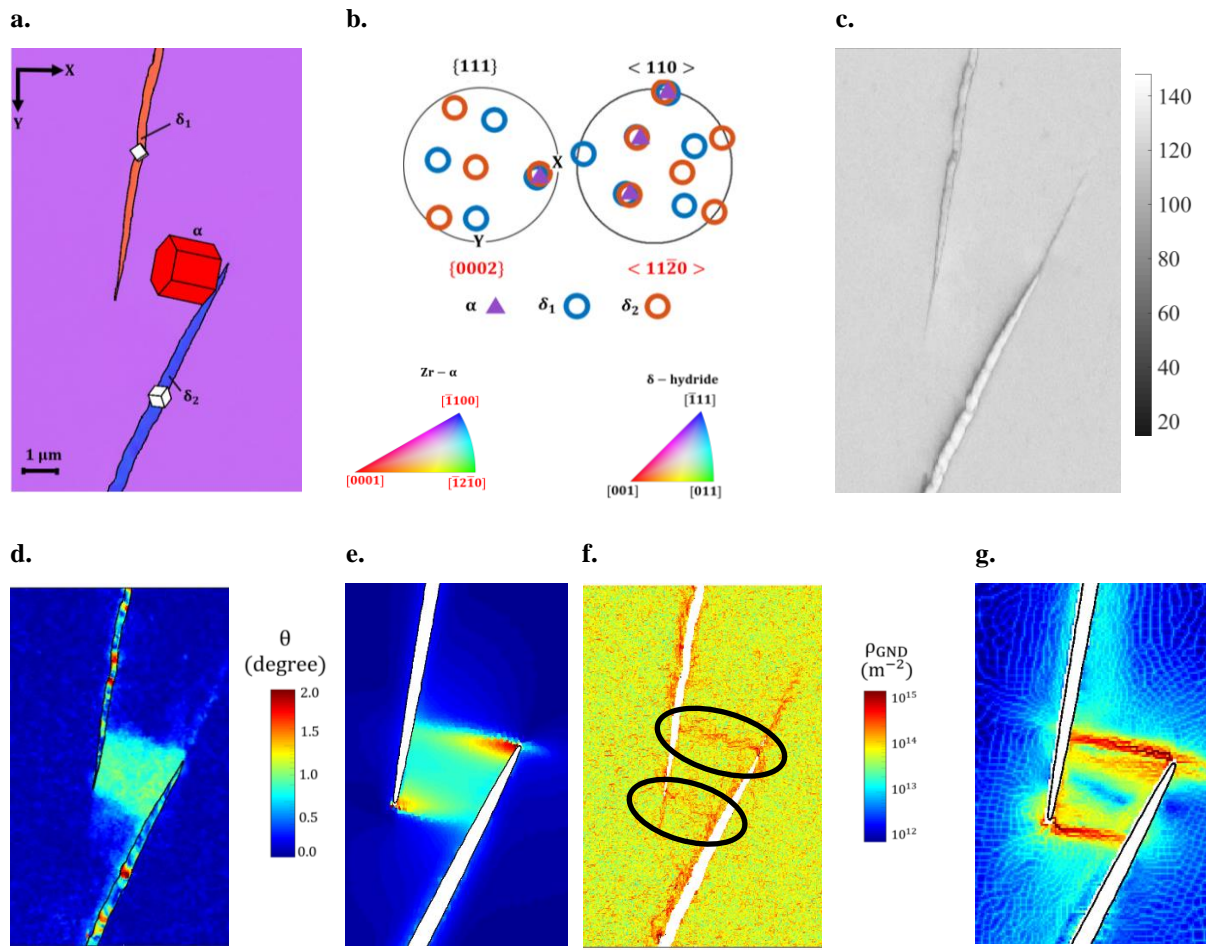

**Fig. 3.** The interaction of hydrides: (a) EBSD orientation map, (b) the orientation relationship between the α parent grain and the two δ-hydrides. Orientations in (a) are with respect to the inverse pole figure Z, with the legends shown in (b). (c) As-measured band contrast map. The misorientation map from (d) EBSD and (e) CPFE. Geometrically necessary dislocation densities from (f) EBSD and (g) CPFE.

The results of CPFE modeling for the stress field induced by the formation of hydrides are shown in Fig. 4. The highest stress component is $\sigma_{xx}$ which varies between ±450 MPa. Similar to GROD and GND maps, a polarized stress field is observed in between the hydrides where tensile stresses are observed at the hydride tips, and compressive stresses are observed in between the two hydrides. These variations are provided in Fig. 4e and along Path #1 shown in Fig. 4a, where a clear sign reversal can be seen for all stress components. Further, the total accumulated plastic shear on all slip systems is shown in Fig. 4d, which explains the trends observed for GROD, GND, and stress maps. The accumulated plastic shear is as high as 5% in the vicinity of hydride tips and the boundary of hydrides. This plastic shear immediately diminishes with distancing from hydrides boundaries leading to the formation of the polarized stress and rotation field in between the hydrides. It is worth mentioning that for calculating the accumulated plastic shear strain, the absolute values of slip are used.

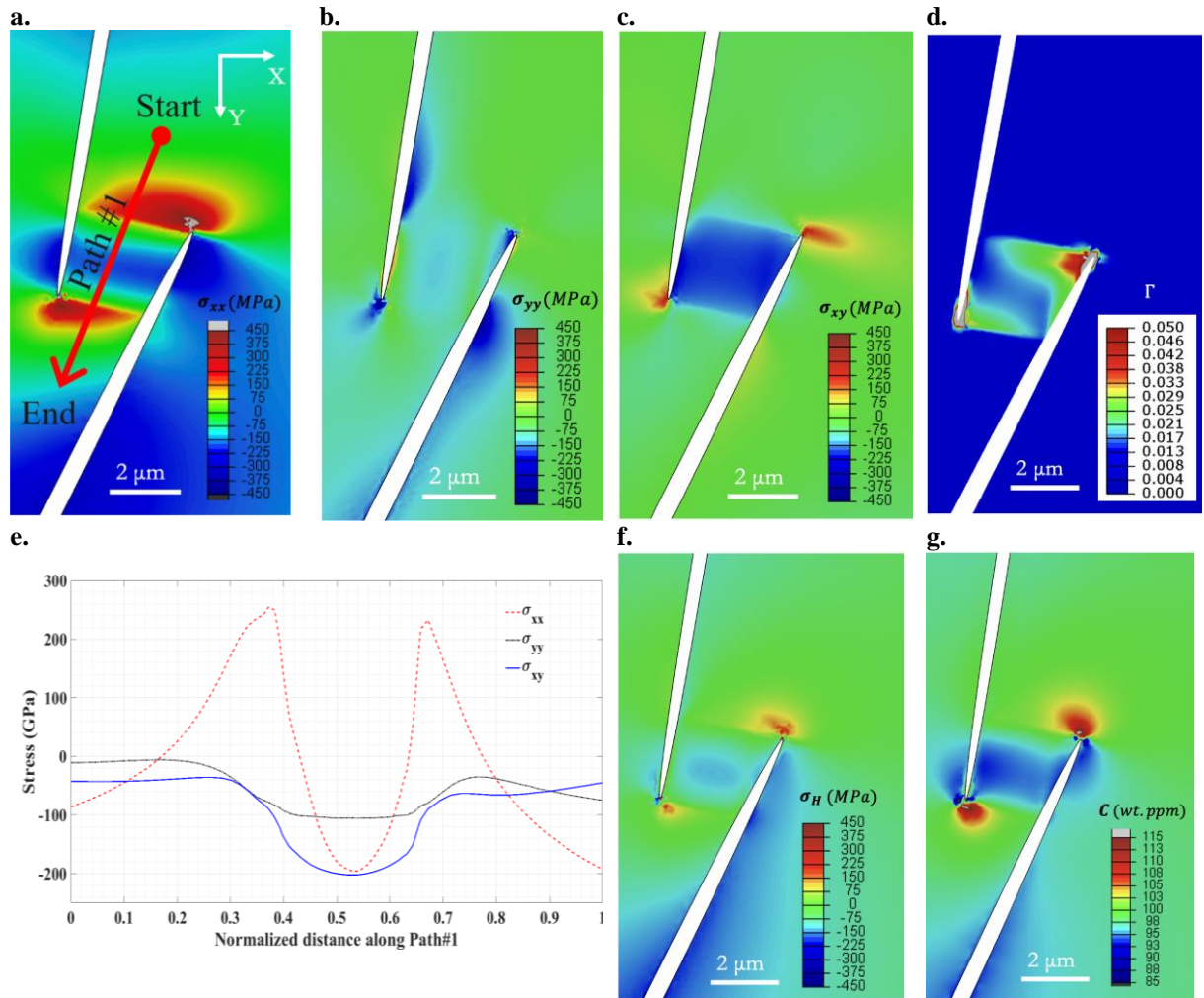

**Fig. 4.** The distribution of (a) $\sigma_{xx}$, (b) $\sigma_{yy}$, (c) $\sigma_{xy}$, and (d) the total plastic shear accomodated by all active slip systems around the intergranular hydrides. (e) The variation of stress components along the path #1 shown in (a). The distribution of (f) hydrostatic stress ($\sigma_H$) and (g) the corresponding hydrogen mass concentration. Results presented are all from CPFE modeling.

The further progression of hydrides depends on the status of hydrogen field around them. The distribution of hydrostatic stress ($\sigma_H$) and the corresponding hydrogen field within the lattice (C) are respectively shown in Fig. 4f and 4g. In agreement with other stress components, a compressive hydrostatic stress field is observed in between the hydrides causing the depletion of hydrogen atoms, hence, the absence of hydride interlinking. This is further reinforced by the high tensile hydrostatic stresses that are developed at the hydride tips leading to the accumulation of hydrogen atoms and further possible elongation of hydrides in their axial direction. The results presented in this section are consistent with those reported by Tondro et al [74,86], for the interaction of hydrides.

### 3.2 Interacting packs of hydrides

The EBSD and CPFE results for a pack of five intragranular hydrides with three distinct orientations and two parent grains are shown in Fig. 5. The large α-grain that has its c-axis oriented nearly towards the y-axis is called $\alpha_1$ (see Fig. 5a). The orientation relationship analysis shown in Fig. 5b indicates that $\alpha_1$ is the parent grain of two of the hydrides, $\delta_1$ and $\delta_2$. As shown in the magnified region of Fig. 5a, a very thin twin is observed in the parent $\alpha_1$ grain. This twin is called $\alpha_2$ and is the parent grain of hydride $\delta_3$. The as-measured band contrast map is shown in Fig. 5c.

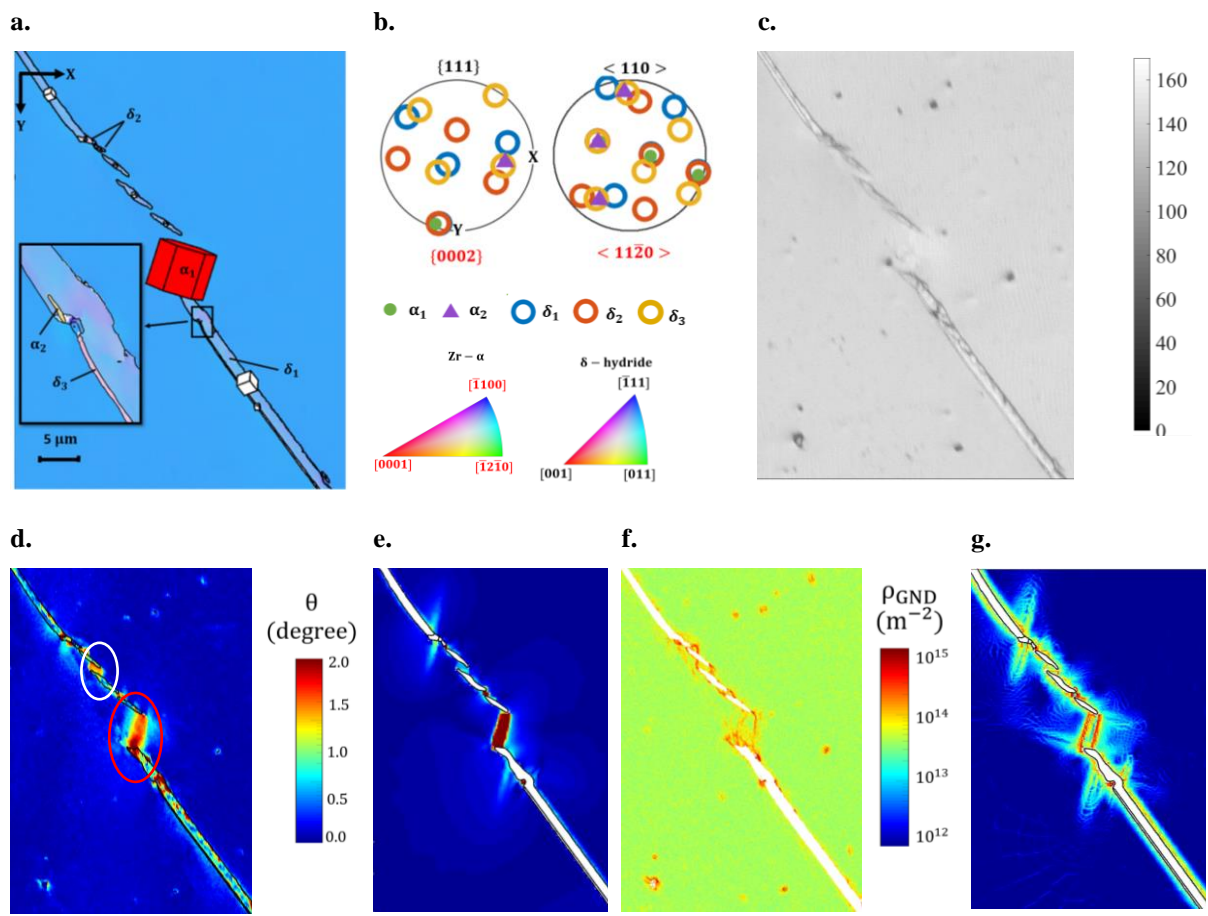

**Fig. 5.** The interaction of hydride packs: (a) EBSD orientation map, (b) the orientation relationship between the α parent grains and the three δ-hydrides. Orientations in (a) are with respect to the inverse pole figure Z, with the legends shown in (b). (c) As-measured band contrast map. The misorientation map from (d) EBSD and (e) CPFE. Geometrically necessary dislocation densities from (f) EBSD and (g) CPFE.

The measured and calculated GRODs for the hydride pack are respectively shown in Fig. 5d and 5e. A distinct rotation field is observed in between neighboring hydrides indicating a strong interaction. Two very localized rotation fields are observed in between the hydrides. For the three smaller $\delta_2$ hydrides, these two rotation fields connect one tip of a hydride to the middle of another hydride. A white oval is plotted around one of these rotation fields in Fig. 5d. The area within the red oval of Fig. 5d shows that the two rotation fields are nearly merged into one, when one of the bottom $\delta_2$ hydride interacts with the bigger hydride $\delta_1$.

The subsequent effects of rotation fields can be seen in the GND maps of Figs. 5f and 5g. A very good agreement is observed between the CPFE and EBSD results. GNDs in both cases increase from $10^{12}$ m$^{-2}$ in "far-field" zones to $10^{15}$ m$^{-2}$ in the vicinity of hydrides. The two localized rotation fields that are highlighted with the red and white ovals led to the formation of two distinct GND fields that are observed in both CPFE and EBSD results. The formation of such dislocation fields leads to further trapping of hydrogen atoms, which in combination with hydrostatic stresses may help further axial growth of hydrides. In addition, higher GND and GRODs are measured and calculated at hydride tips in comparison to hydride sides.

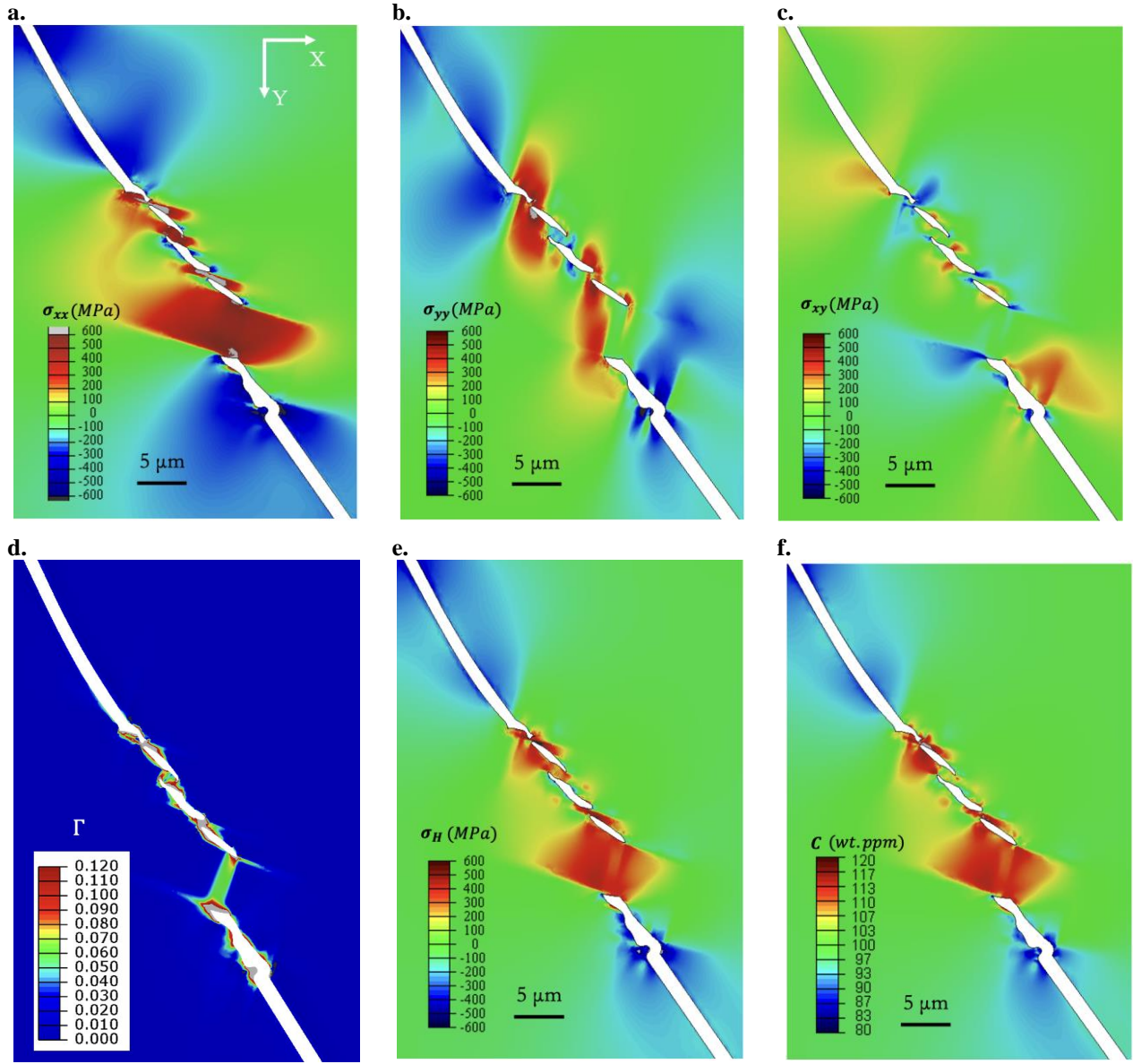

Fig. 6. The distribution of (a) $\sigma_{xx}$, (b) $\sigma_{yy}$, (c) $\sigma_{xy}$, and (d) the total plastic shear accomodated by all active slip systems around the pack of intragranular hydrides. The distribution of (e) hydrostatic stress ($\sigma_H$) and (f) the corresponding hydrogen mass concentration. Results presented are all from CPFE modeling.

The results of CPFE modeling for the stress fields around the pack of interacting hydrides are shown in Fig. 6. Both $\sigma_{xx}$ and $\sigma_{yy}$ are considerably high where a polarized stress field is observed for all stress components. This is due to the mechanism by which α$_1$ parent grain accommodates the localized strain fields induced by the precipitation of hydrides. It is evident from Fig. 6d that plastic strains are localized in between the small areas in between the hydrides. Similarly, a polarized hydrostatic stress field is observed in the vicinity of hydrides. Although, unlike the intragranular hydrides of the previous section, only tensile hydrostatic stresses are observed in between the pack of hydrides. This highlights the significant effects of local microstructure as well as the crystallographic orientations of hydrides and the parent grains on the development of stress patterns. The corresponding hydrogen field is shown in Fig. 6f, and it is observed that hydrogen atoms are concentrated both in between the hydrides as well as hydride tips. Due to the size and shape of hydride packs, and due to the strong interactions among them, the concentration of hydrogen can be as high as 120 wt.ppm, which is much higher than the one calculated for the two intergranular hydrides of previous section.

### 3.3 Grain boundary hydrides

The deformation induced by hydrides in two neighboring α-grains is studied in this section. For this purpose, a large hydrided grain boundary was selected, but divided into three subsections that are called Regions of Interest (ROIs). All three ROIs and the hydrided grain boundary are shown in Fig. 7a. Considering the size of hydrides and parent grains, high spatial resolution EBSD measurements were conducted in the ROIs which are shown in the right-hand column of Fig. 7. The c-axis of $\alpha_1$ parent grain is at ~30° angle from the y-axis while the c-axis of $\alpha_2$ parent grain is nearly parallel to the x-axis. The naming of each grain is provided within the HCP schematics of Fig. 7. To avoid busy figures, the poles of $\alpha_1$ and $\alpha_2$ as well as their corresponding hydrides are shown in two different pole figures of Fig. 7c. Four hydrides are observed to form at this grain boundary which are named $\delta_1$ to $\delta_4$. The top pole figure shows the orientations of $\delta_1$ and $\delta_2$, while the bottom pole figure shows the orientations of $\delta_3$ and $\delta_4$. The orientation relationship analysis presented in Fig. 7c indicates that $\alpha_1$ is the parent grain of $\delta_1$ and $\delta_2$ hydrides, while $\alpha_2$ is the parent grain of $\delta_3$ and $\delta_4$ hydrides. The two hydrides $\delta_3$ and $\delta_4$ are visible in the EBSD map of ROI$_1$, whereas $\delta_1$ and $\delta_2$ hydrides are visible in ROI$_2$. Although the hydride map is divided into three ROIs, CPFE modeling was conducted on all connected hydrides, but for a like-to-like comparison and to study variations, numerical results for GND and rotation maps are also sectioned to three ROIs. The stress maps are provided for the entire CPFE model, though.

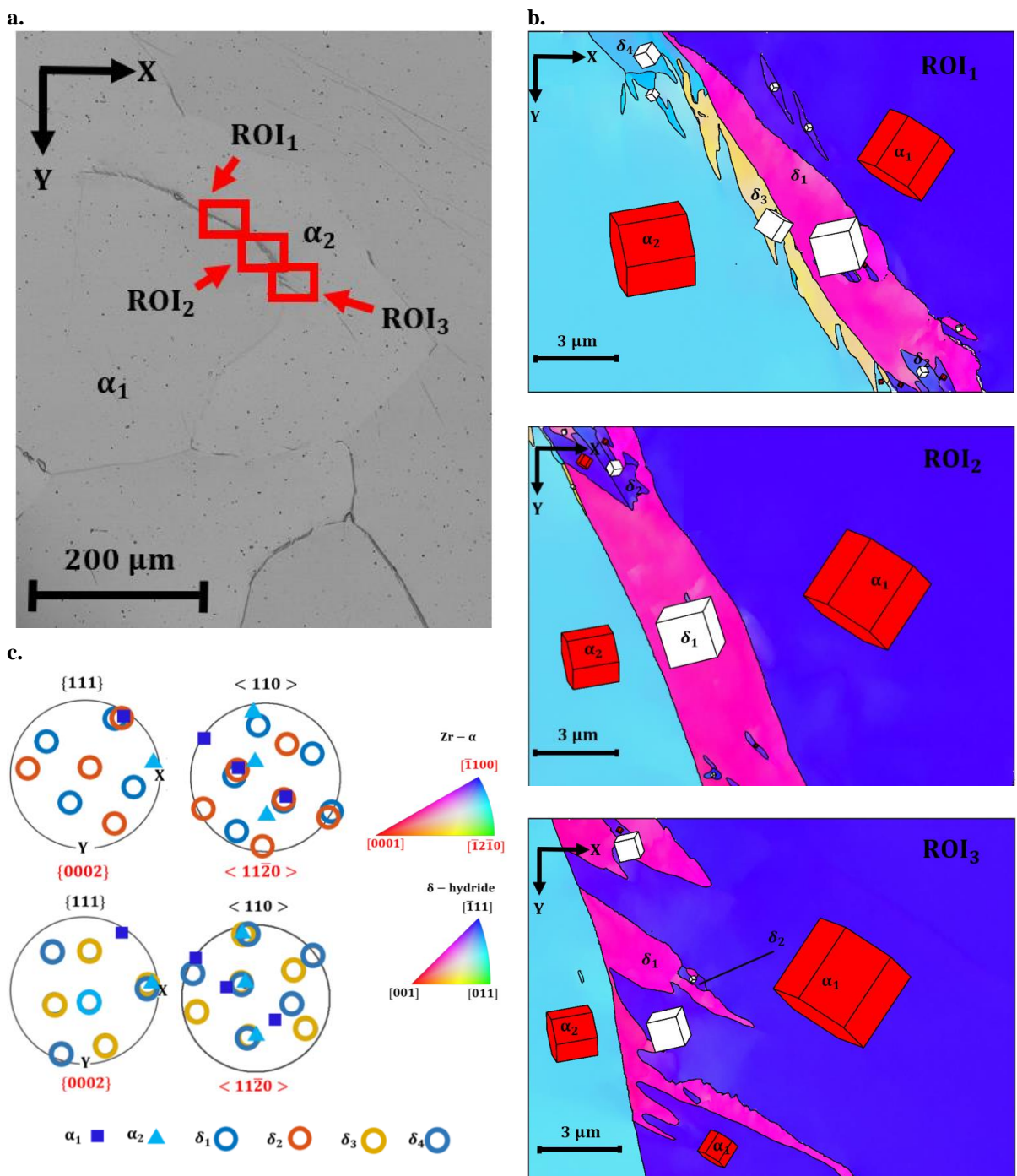

**Fig. 7. (a) An SEM image of the investigated hydrides that are located at the grain boundary of two α-grains. This area is divided into three Regions of Interests (ROIs), which are shown in the second column. Four distinct hydrides are observed and their orientation relationship with the neighbouring α-grains are shown in (c).**

The measured and calculated GRODs and GNDs for the three ROIs are shown in Fig. 8. The results for $ROI_1$, $ROI_2$ and $ROI_3$ are respectively shown in the first, second, and third columns. The misorientation maps are shown in the first two rows while GNDs are shown in the last two. Both CPFE and EBSD results reveal a high lattice rotation in the vicinity of hydrides, particularly at hydride tips. The tips of hydrides are mainly seen in $ROI_1$ and $ROI_2$ (Fig. 8a, 8c, 8d and 8f) where misorientations exceed 4°, which is twice the magnitude measured for intragranular hydrides. The rotation fields are very localized and are highlighted by the ovals in Fig. 8b and 8c.

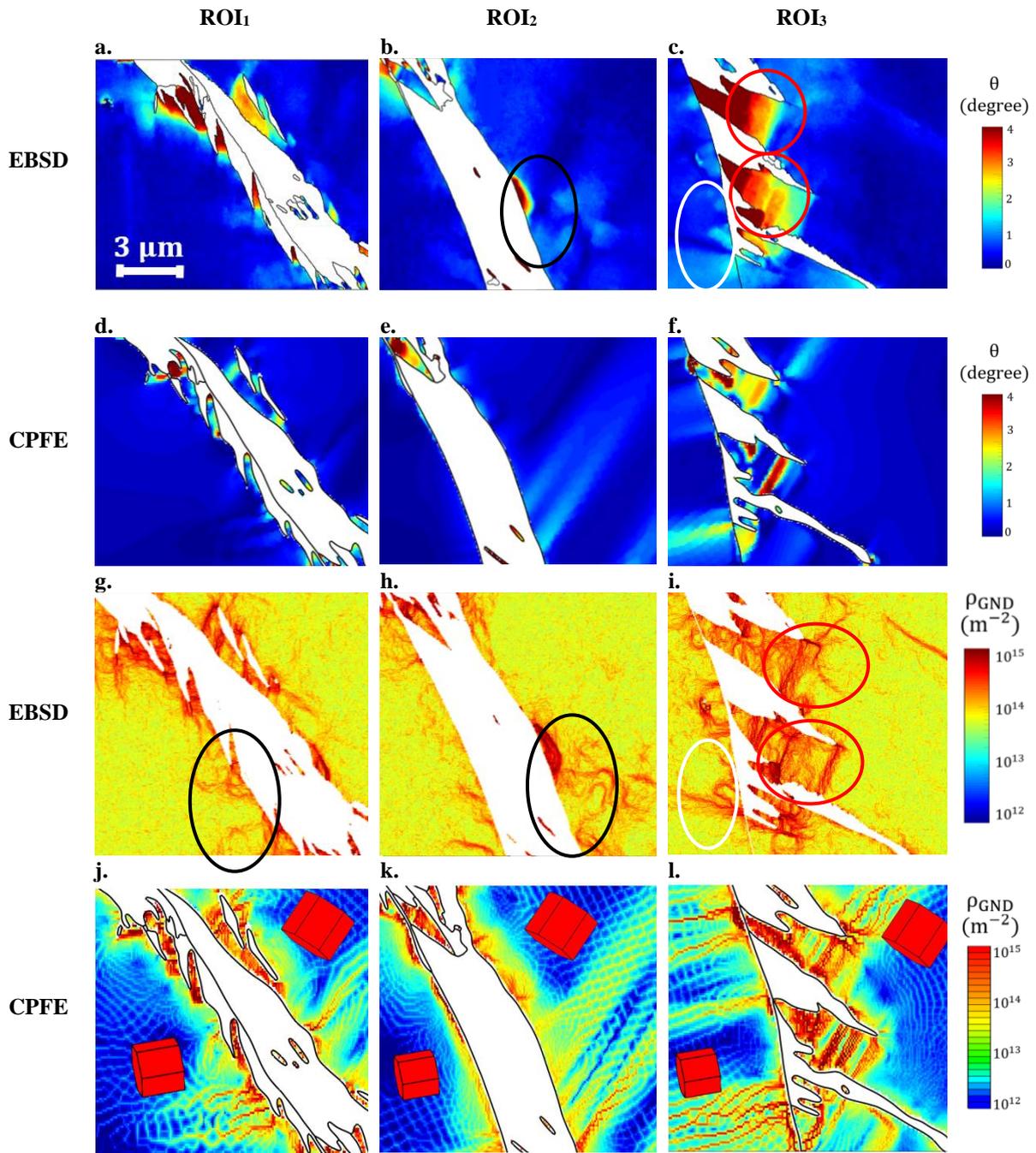

**Fig. 8. The distribution of misorientations and geometrically necessary dislocation (GND) densities within α parent grains from EBSD and CPFE. The first and second rows represent the misorientation maps from EBSD and CPFE, respectively. The third and fourth rows respectively represent the GND maps from EBSD and CPFE. The results for the Region of Interests (ROI) are respectively shown in each column. ROIs are shown in Fig. 7.**

The subsequent effects of misorientation fields can be seen in the GND maps of Fig. 8. An outstanding agreement is observed between CPFE and EBSD results, further highlighting the significant effects of hydride transformation strain on inducing localized deformation fields.

While a high GND concentration is observed in the vicinity of hydrides, discrete parallel GND densities are observed in all ROIs. For example, the red ovals of Fig. 8i highlight parallel and discrete GNDs, both in CPFE and EBSD results. The GNDs are mainly parallel to the c-axis of the α-grain, and in some cases are parallel to the basal slip system, indicating the high slip activity in the $α_1$ parent grain. Likewise, the white oval of Fig. 8i points to the formation of localized GND field that is parallel to c-axis of the $α_2$ parent grain. It is interesting to see the formation of discrete GNDs both in CPFE and EBSD results of Figs 8i and 8l. addition, the black oval of Fig. 8h highlights a localized GND field induced by the formation of $δ_1$ within $α_1$. CPFE simulation shows that mainly basal and to a lesser extent prism slip systems are active leading to the formation of such localized GND zones (Fig. 8k and Fig. 11).

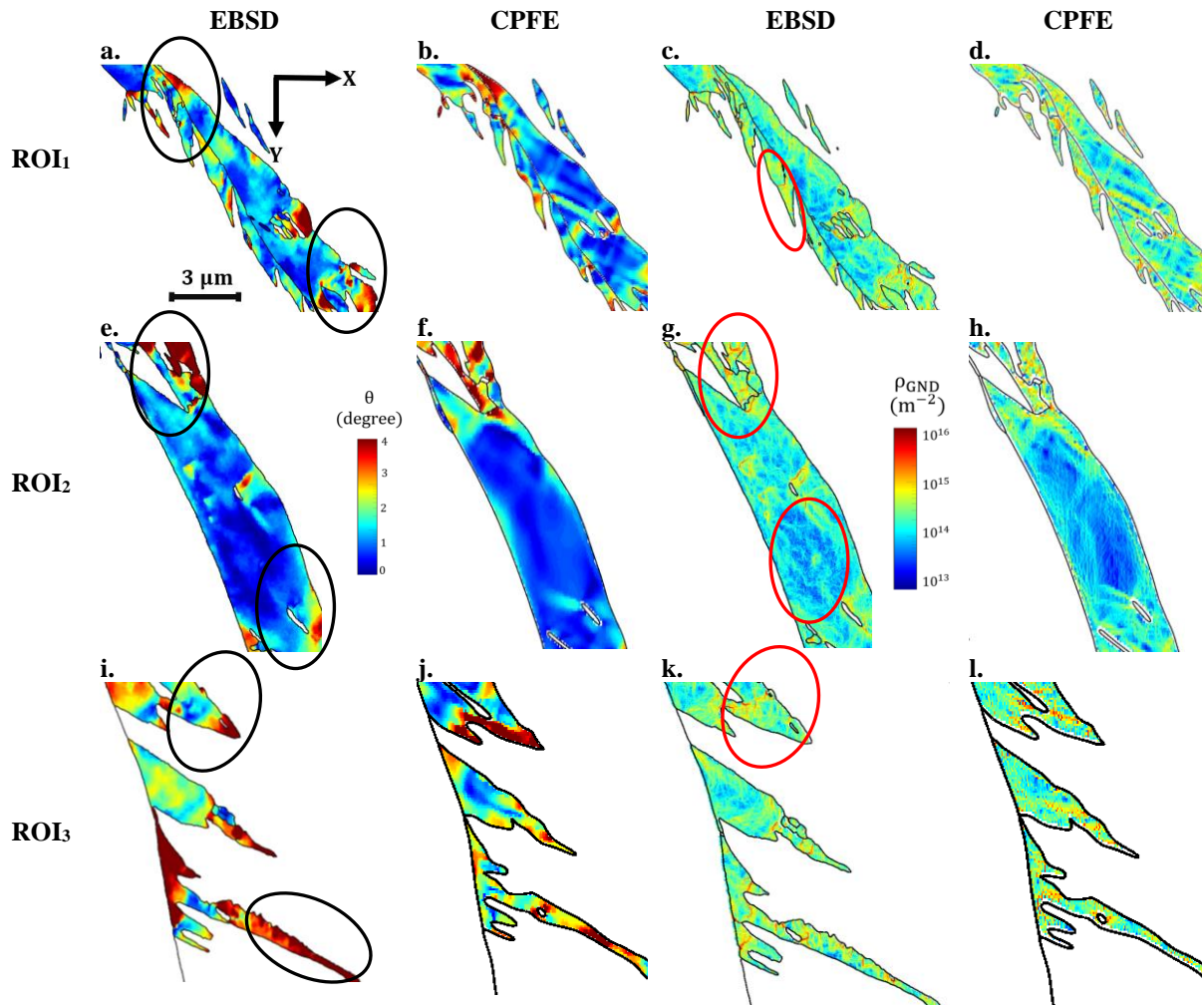

**Fig. 9. The distribution of misorientations and geometrically necessary dislocation (GND) densities within δ hydrides from EBSD and CPFE. The first and third columns represent EBSD results for misorientations and GNDs, while the second and fourth columns represent the corresponding CPFE results. The results for the Regions of Interest (ROI) are respectively shown in each row. ROIs are shown in Fig. 7. The legend for misorientations and GNDs are respectively shown in 9e and 9g.**

So far, variations within the α parent grains are shown, with less attention paid to the thin hydrides, however, the hydrides of this section are thick enough that allows studying misorientation and GND variations across them. In particular, the hydrides of $ROI_1$ and $ROI_2$, as shown in Fig. 9, are thicker than 3 µm, allowing further collections of data points. In Fig. 9a and 9b, it is shown that while the tips of hydrides have high GRODs, the middle section of hydrides undergo less GRODs suggesting that hydride tips are highly stressed in comparison to the middle sections. The high GROD zones are highlighted with the black ovals of Fig. 9. Likewise, low GND densities are measured and calculated in the middle section of hydrides while higher GND densities are observed at the hydride tips. For example, the red oval of Fig. 9c points to a hydride tip that has a high GND density both in CPFE and EBSD results. On the other hand, a smaller GND density is observed in the middle of hydride next to the same red oval. The same trend is observed in Fig. 9g and 9h, where a clear dislocation "free zone" is observed in the middle of hydride. Consistent with other measurements, the black and red ovals of Fig. 9i to 9l reveal high GROD and GND densities in at the hydride tips, within the hydride itself.

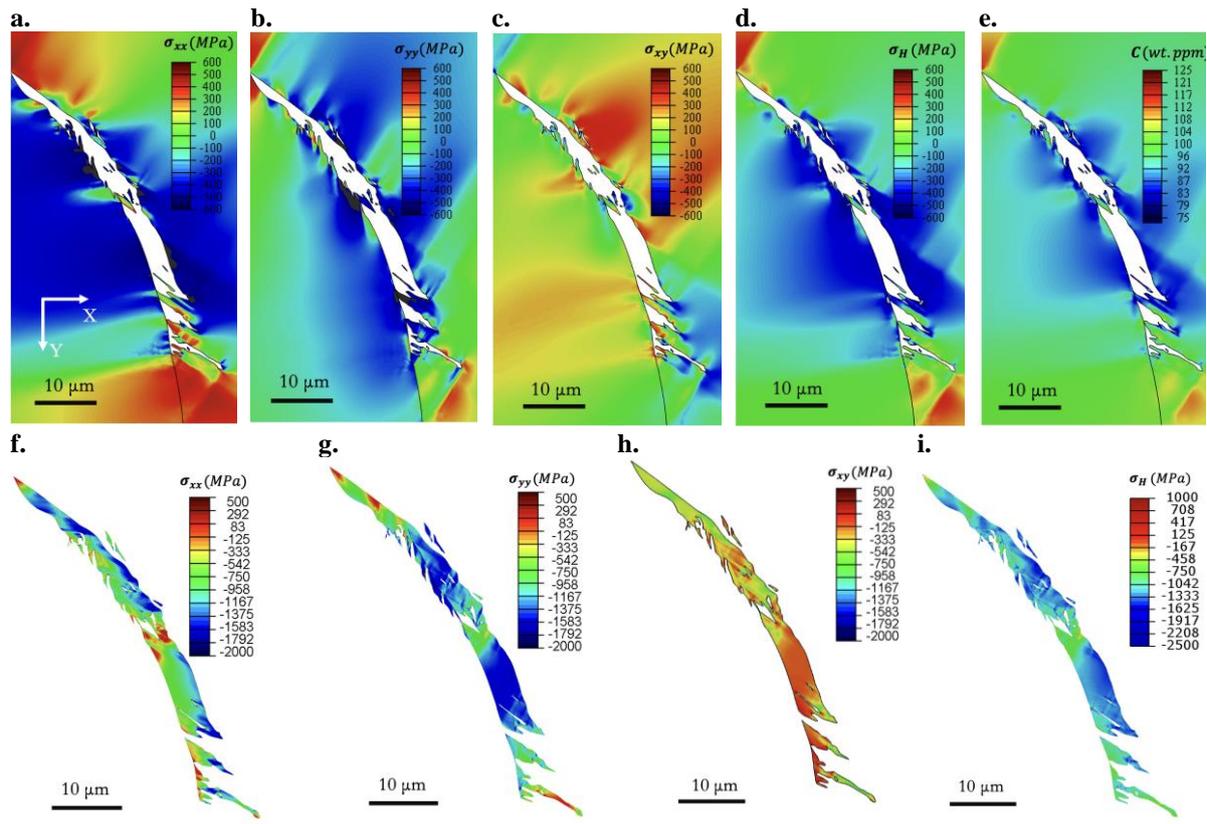

**Fig. 10. The distribution of (a) and (f) $\sigma_{xx}$, (b) and (g) $\sigma_{yy}$, (c) and (h) $\sigma_{xy}$, (d) and (i) hydrostatic stress ($\sigma_H$), and (e) hydorgen mass concentration (wt. ppm). The CPFE results for α parent grains and hydrides are respectively shown in the first and second row.**

The CPFE results for stress fields around and within the hydrides of all ROIs are shown in Fig. 10. For the α parent grains, the highest components of stress tensor are $\sigma_{xx}$ and $\sigma_{yy}$ which are very negative along the hydride boundaries but are positive at the hydride tips. Similarly, the calculated hydrostatic stress is compressive along the hydride boundaries (Fig. 10d), but it is tensile at hydride tips leading to respectively depletion and accumulation of hydrogen atoms at the mentioned locations (Fig. 10e). This trend holds for both α parent grains, despite having different crystallographic orientation, and that four different δ hydrides were precipitated in them. In addition, consistent with the GROD and GND calculations, stress concentrations are observed right at the hydride boundary. Hydrides are removed from the first row for better visualization and to clearly see the stress variations within the α parent grains. The stress fields within hydrides are shown in the second row of Fig. 10. Compressive stresses as high as -1

GPa are calculated within hydrides, which is in agreement with the dislocation dynamic calculations of Tummala et al [56].

The calculated total accumulated plastic shear on all slip systems of α parent grains are shown in Fig. 11. Consistent with the GROD and GND maps, a high plastic shear is observed in the vicinity of hydrides, particularly at the hydride tips. The active slip systems within the parent grains are shown by the blue arrow and highlighted slip plane. It is evident that the basal slip system is very active in both α parent grains.

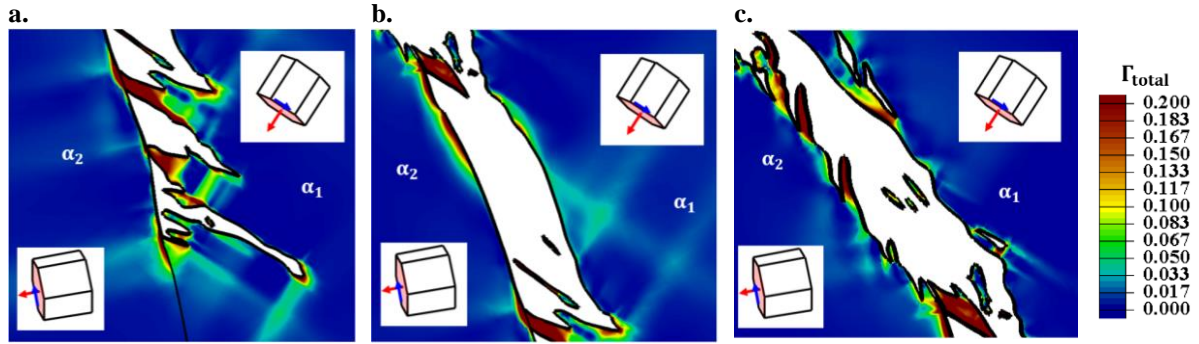

**Fig. 11. The distribution of total plastic shear strain accumulated all active slip systems within α parent grains: (a) ROI$_1$, (b) ROI$_2$, (c) ROI$_3$.**

### 3.4 Triple point hydrides

In this section, CPFE and EBSD results for the formation of hydrides at a triple point are provided. Fig. 12a shows an SEM image of the investigated area, where three α parent grains and four distinct hydrides were observed. The measured EBSD map of parent grains and hydrides are shown in Fig. 12b. A rather thick hydride was formed at the grain boundary shared between α$_1$ and α$_2$, which are located at the bottom of Fig 12b. This hydride is called δ$_1$, and the orientation analysis shown in Fig. 12d revealed that α$_2$ is its parent grain. Another hydride was observed at the right corner of the triple point, which is called δ$_2$. As shown in the orientation analysis, δ$_2$ hydride was formed inside the α$_2$ parent grain. The third hydride was observed in the left corner of the triple point, called δ$_3$, and has an orientation relationship with α$_3$ parent grain. The last hydride was observed within α$_3$ and is called δ$_4$. This hydride also follows the hydride orientation relationship with its parent grain, α$_3$.

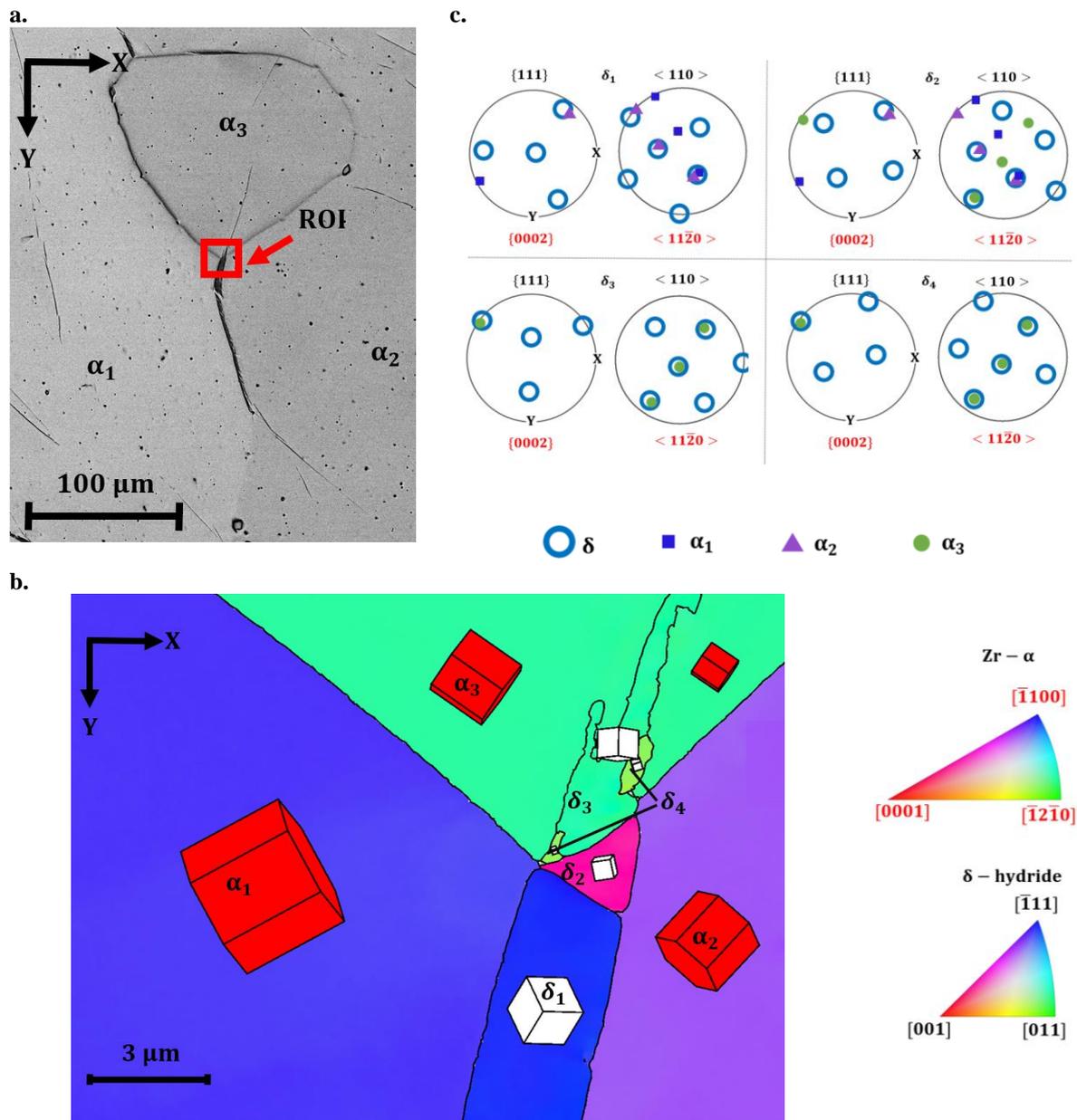

**Fig. 12.** (a) An SEM image of the investigated hydrides that are located at a triple point. (b) The crystallographic orientation of α parent grains and hydrides. Three α grains and four distinct hydrides are indexed. (c) The crystallographic relationship between each indexed hydride and the corresponding α parent grain. Hydrides are shown with blue circles.

The measured and calculated GROD, as well as GND densities for both α parent grains and four δ hydrides are shown in Fig. 13a- 13h. A very good agreement is observed between CPFE and EBSD results for both GRODs and GND maps. A very high GROD is measured by EBSD within $α_1$ which is highlighted by the bigger black oval of Fig. 13a. Although none of the hydrides had an orientation relationship with $α_1$, the transformation strain induced a large rotation within this grain which can also be seen in the CPFE results of Fig. 13b. This large rotation resulted in the formation of discrete GND zones that are observed in both EBSD and CPFE results for the same grain (see Fig. 13e and 13f). The total accumulated plastic shears on all slip systems of α grains are shown in Fig. 13i. It is evident from CPFE results that the observed high GROD and GND fields within $α_1$ is a result of accumulation of plastic shear in the same location. Likewise, a high GROD field is observed in $\alpha_2$ parent grain which is highlighted with the smaller black oval of Fig. 13a. A high intensity GROD field is also observed in the CPFE results at the same location (Fig. 13b). The subsequent of this rotation field can be seen in the GND maps of Fig 13e and 13f, where discrete parallel GND fields are observed in both experimental and modeling results. The calculated plastic shear presented in Fig. 13i shows that basal slip system is very active followed by prism. Similar trends are also observed in $\alpha_3$ parent grain, in the areas highlighted by the black circle of Fig. 13a. A high GND density is calculated in both CPFE and EBSD result.

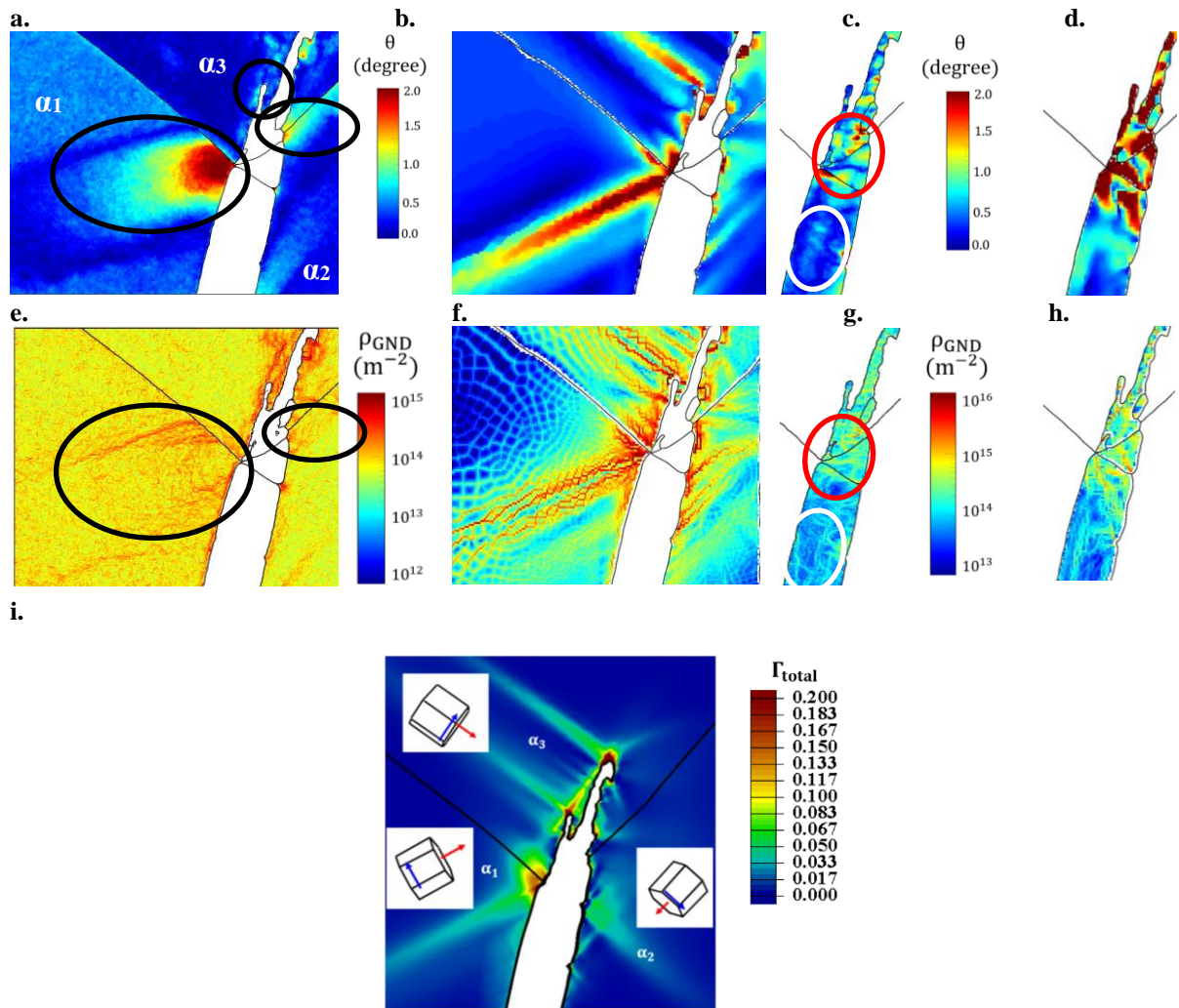

**Fig. 13.** The distribution of misorientations and geometrically necessary dislocation (GND) densities within α parent grains and δ hydrides from EBSD and CPFE. Measured misorientations within (a) α parent grains and (c) δ-hydrides with the corresponding calculated values respectively shown in (b) and (d). Measured GND densities within (e) α parent grains and (g) δ-hydrides with the corresponding calculated values respectively shown in (f) and (h). The total calculated plastic shear on all slip systems for (i) α parent grains

The measured and calculated GRODs for the four δ hydrides are respectively shown in Fig. 13c and 13d. Similar to the results shown in section 3.3 for grain boundary hydrides, higher GRODs are observed in the vicinity of hydride boundaries. An example of such a high GROD zone is highlighted with the red oval of Fig. 13c, which is also observed in the CPFE results of Fig. 13d. Both experimental and modeling results indicate that the minimum variation of GRODs is in the middle of the thicker hydride- which is highlighted with the white oval of Fig.

13c. In agreement with the observed trends for GRODs, higher and lower GND zones are respectively seen in the red and white ovals of Fig. 13g, which are replicated by the CPFE model.

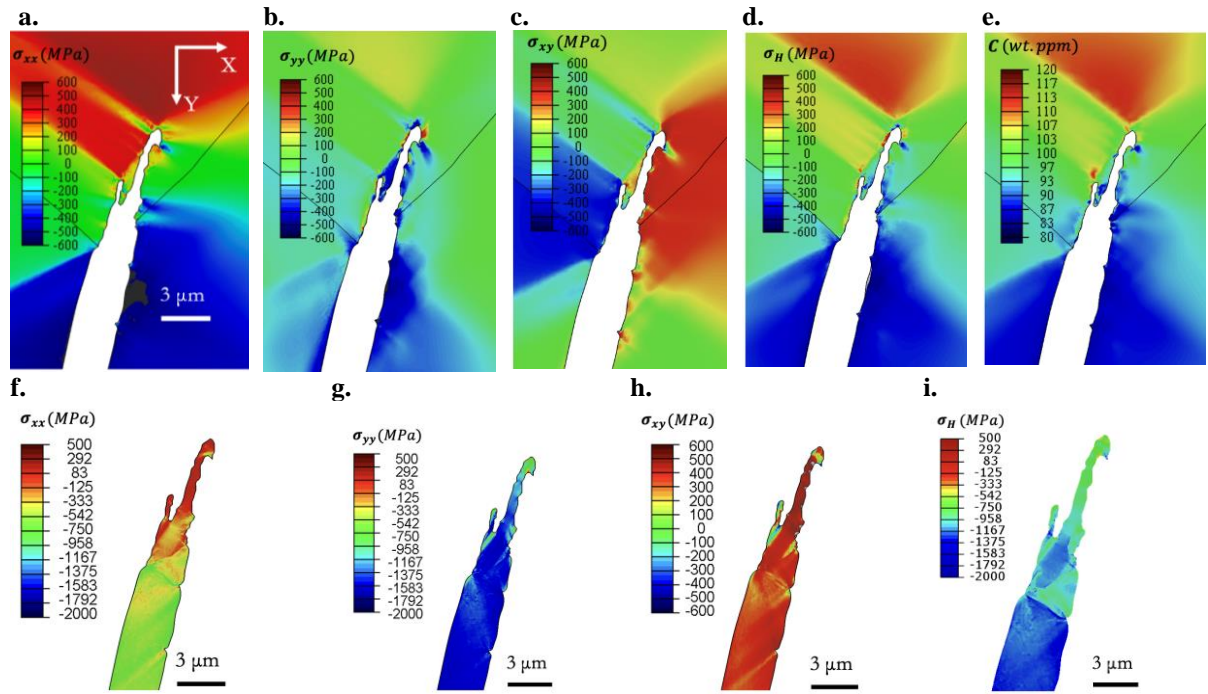

**Fig. 14.** The distribution of (a) and (f) $\sigma_{xx}$, (b) and (g) $\sigma_{yy}$, (c) and (h) $\sigma_{xy}$, (d) and (i) hydrostatic stress ($\sigma_H$), and (e) hydorgen mass concentration (wt. ppm). The CPFE results for α parent grains and hydrides are respectively shown in the first and second row.

The calculated stresses for the three α-grains and hydrides are shown in Fig. 14. In α-grains, the highest component of stress tensor is $\sigma_{xx}$ which is negative in the two bottom grains, $\alpha_1$ and $\alpha_2$, while it is tensile and positive in the top grain $\alpha_3$. The corresponding hydrostatic and mass concentration fields are respectively shown in Fig. 14d and 14e, where similar polarized fields are observed. The depletion of hydrogen atoms from the sides of $\delta_1$ hydride, located in the bottom of the figure, and further accumulation of hydrogen atoms at the tips of $\delta_3$ hydride located in the top α grain, might be reason why the hydride penetrated the top grain and not further expanded in the two neighboring $\alpha_1$ and $\alpha_2$ grains. Consistent with the results of the previous section, the two normal components of stress tensor within hydrides are very negative, leading to a very negative hydrostatic stress.

## 4. Discussion

In this section, the parameters that contribute to the results of CPFE modeling are discussed first. This is followed by a detailed investigation of how surrounding α-grains accommodate the plastic deformation induced by hydride formation, and how GNDs vary with distancing from hydride-zirconium interfaces. Lastly, the effects of thermal cycling on the evolution of GNDs and stress fields are studied.

### 4.1 The mechanical properties of hydrides

The effects of the mechanical response of hydrides are studied by assigning different CRSS to the hydride domain so that an upper and lower bound for the calculated misorientation fields can be obtained. This is done by using fully elastic properties and lower CRSS values to cover a wide spectrum for hydride's plastic properties. Please note that the determination of bulk hydride's mechanical properties is not in the scope of this section. Indeed, the goal of this section is to determine if the conclusions made from CPFE modeling depend on the mechanical properties of the hydride domain and whether the observed trends for rotation fields are affected.

The results of CPFE modeling for the intragranular hydrides of section 3.1 are revisited here by changing hydride's CRSS values from 220 MPa, resembling a soft hydride, to infinity that resembles a fully elastic hydride. CPFE results are shown in Fig. 15a to 15e, while the EBSD results for the same map is shown in Fig. 15f. Misorientations are concentrated in between the hydrides when CRSS varies between 220 MPa to 460 MPa, indicating that trends reported are not affected. When plastic deformation inside hydride is switched off, a high misorientation field is observed in between the hydrides as well as along the hydride boundaries, which are significantly stronger than what is observed in the experiment. This indicates that hydrides may not be fully elastic. This observation is in agreement with several studies focusing on the deformability of hydrides in Zr and Ti alloys [9,36–39,41,42,44–46]. The experimental and modeling results of this section also indicate that hydride transformation strain is partially accommodated by the plastic deformation inside hydrides. This observation is also in agreement with previous studies indicating that the transformation strain can induce plasticity both internally and at the surface of hydrides [87,88].

Further, in Fig. 15g, CPFE and EBSD results for misorientations along Path #1, shown in Fig 15f, are plotted. The modeling results for CRSS of 220 MPa, 340 MPa, and 460 MPa follow those from EBSD, while those from 800 MPa, 2000 MPa deviate. The highest deviation from experimental results is observed for the elastic hydride further indicating that hydrides may not be elastic. In addition, the hydride CRSS used in the result section was taken from recent micropillar experiments [24], which shows a good agreement with EBSD results.

CPFE results with different CRSS values for grain boundary hydrides of section 3.3 are provided in the supplementary Fig. S2. Similar trends to the ones stated in the previous paragraph are observed further reinforcing the conclusions made regarding the misorientation fields around the hydrides.

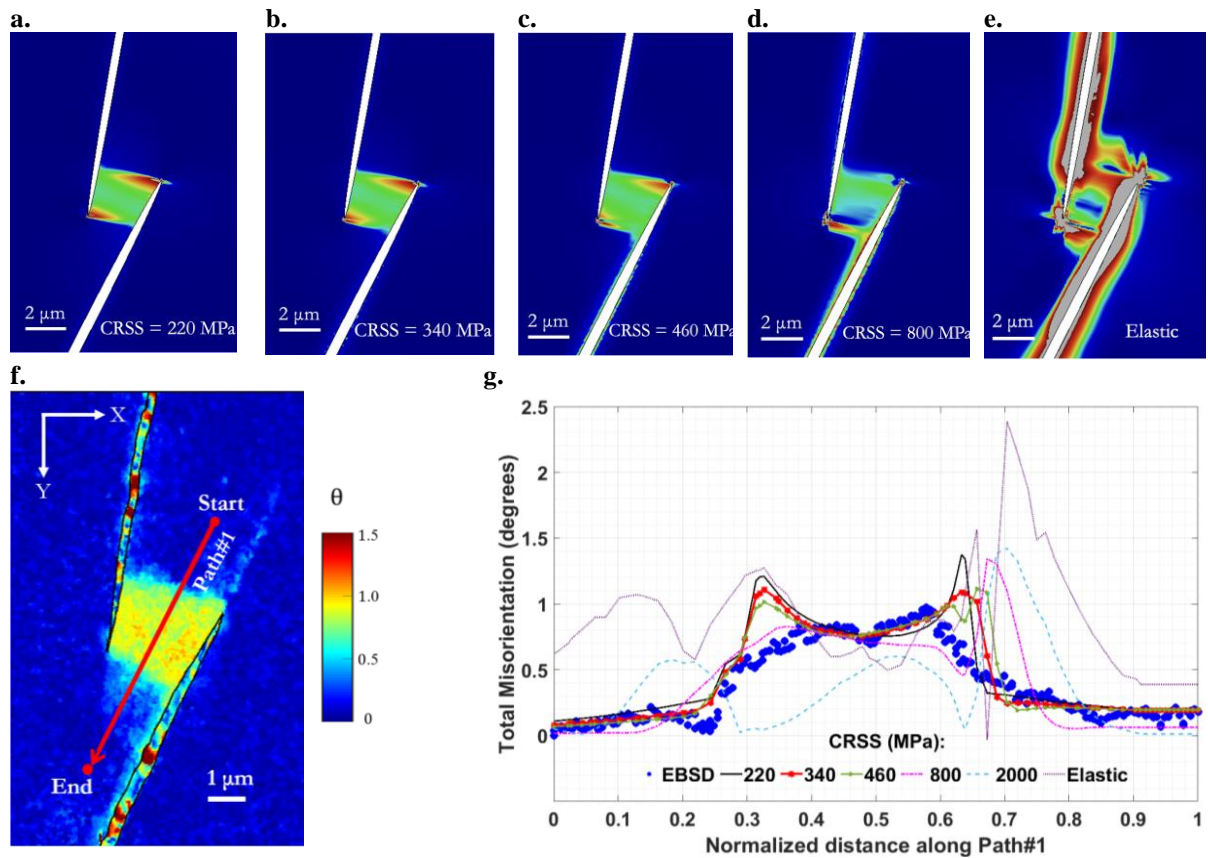

**Fig. 15.** The effects of hydrides critical resolved shear stress (CRSS) variations on the calculated misorientation fields from CPFE: CRSS of (a) 220 MPa, (b) 340 MPa, (c) 460 MPa, (d) 800 MPa, and (e) fully elastic. (f) The measured misorientation field from EBSD for two interacting intragranular hydrides. (g) A comparison between the results of CPFE modeling with different CRSS values for hydrides with those from EBSD, along the Path #1 shown in (f).

### 4.2 The sequence of hydride formation

So far, the transformation strain induced by hydride precipitation has been applied to all hydride domains simultaneously. Here, the effects of precipitation sequence are studied to investigate whether the CPFE results presented are impacted. For this purpose, the pack of hydrides presented in section 3.2 was simulated and results are provided in Fig. 16. For this model, the hydride transformation strain was first applied to the longer hydrides located at the top and bottom of the model while the pack of small hydrides did not undergo any phase transformation. The results for the first step are shown in the first row of Fig. 16. In the second step, transformation strain was applied to the pack of smaller hydrides and results are presented in the second row of Fig. 16.

The CPFE results for the first step show that $\sigma_{xx}$ and $\sigma_{yy}$ are the two major components of stress tensor with a shape that seems to be reflected in the morphologies of smaller hydrides. Interestingly, the pack of smaller hydrides are in the area where higher tensile hydrostatic stresses are calculated (Fig. 16d). The shapes of the stress fields are changed significantly when smaller hydrides are not precipitated yet are almost identical to the ones presented in Fig. 6 for the case where all hydrides precipitated simultaneously. That is, the sequence of hydriding is not a major contributor to the trends presented in this paper.

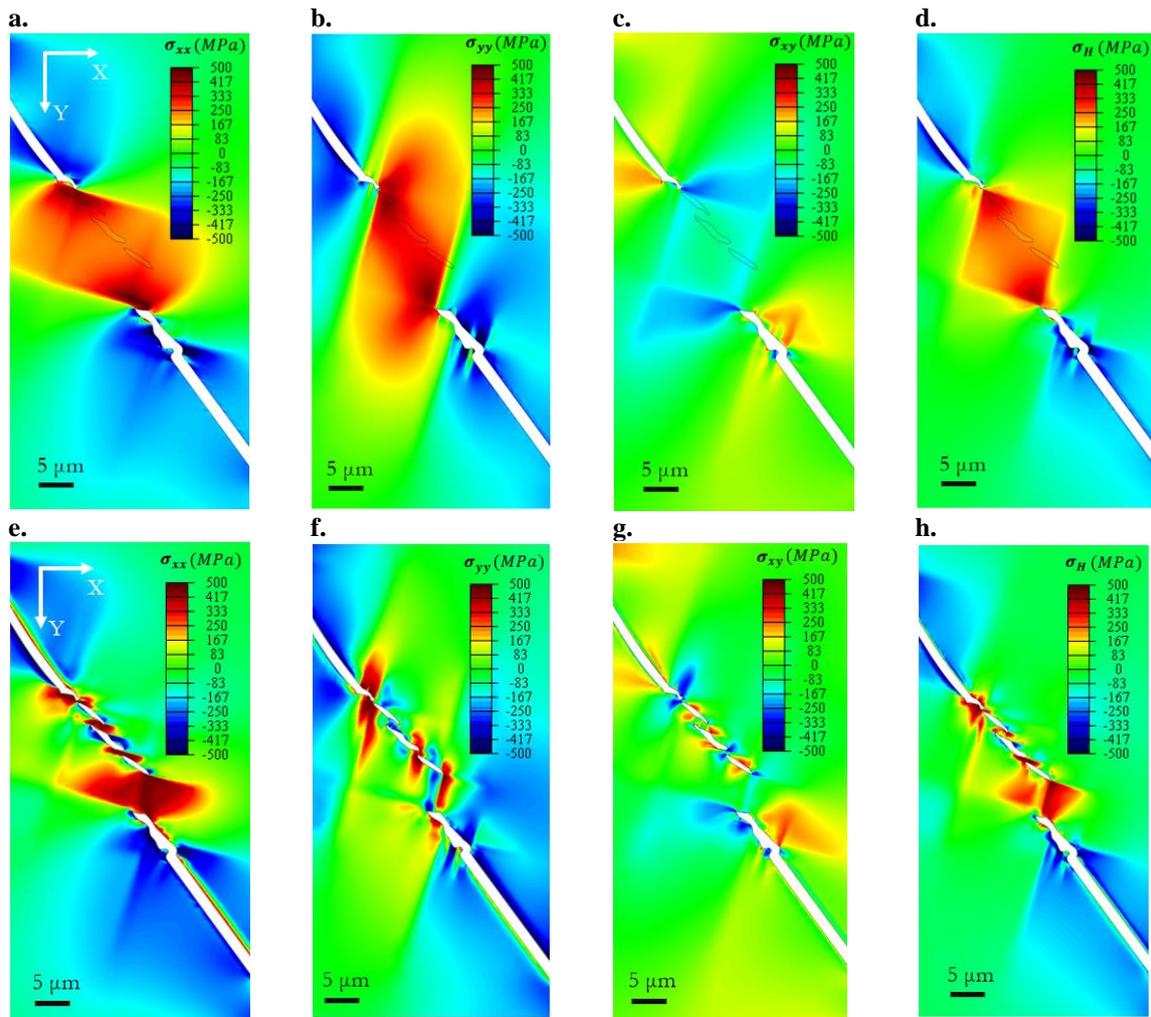

**Fig. 16.** The effects of hydride precipitation sequence. (a)-(d) The stress induced by the precipitation of top and bottom hydrides, and (e)-(h) those when the smaller pack of hydrides are subsequently transformed. CPFE results for (a) and (e) $\sigma_{xx}$, (b) and (f) $\sigma_{yy}$, (c) and (g) $\sigma_{xy}$, (d) and (h) hydrostatic stress $\sigma_H$.

### 4.3 The distribution of GNDs and shear strains around various hydrides

In this section, the spatial distribution of GND densities, cumulative plastic shear on all slip systems, and the activity of various slip systems around hydrides within the neighboring α-grains for the four different hydride cases are studied in detail. This is to investigate the relationship between hydride configurations and the deformation fields induced in the surrounding environment.

Fig. 17 shows the spatial variations of the measured GND densities in the surrounding α-grains with distancing from hydride-zirconium interfaces. Specifically, Fig. 17a and 17b display GND variations for two interacting hydrides and interacting packs of hydrides, respectively, while Fig. 17c and 17d display such variations for intergranular and triple point hydrides, respectively. A grayscale color is used to show the frequency of measured values at each distance while the 10%, 90%, and medians are highlighted with the colored points. For the sake of clarity and further analysis, a function of the form $\rho_{GND} = Ae^{-Bx} + C$ is used to fit the green lines to the measured medians. The fitted parameters are provided in Table 2. While this function is descriptive without explicit physical meaning, it effectively represents the patterning of GND densities near hydrides. The average thickness of hydrides for each category is also provided in Table-2.

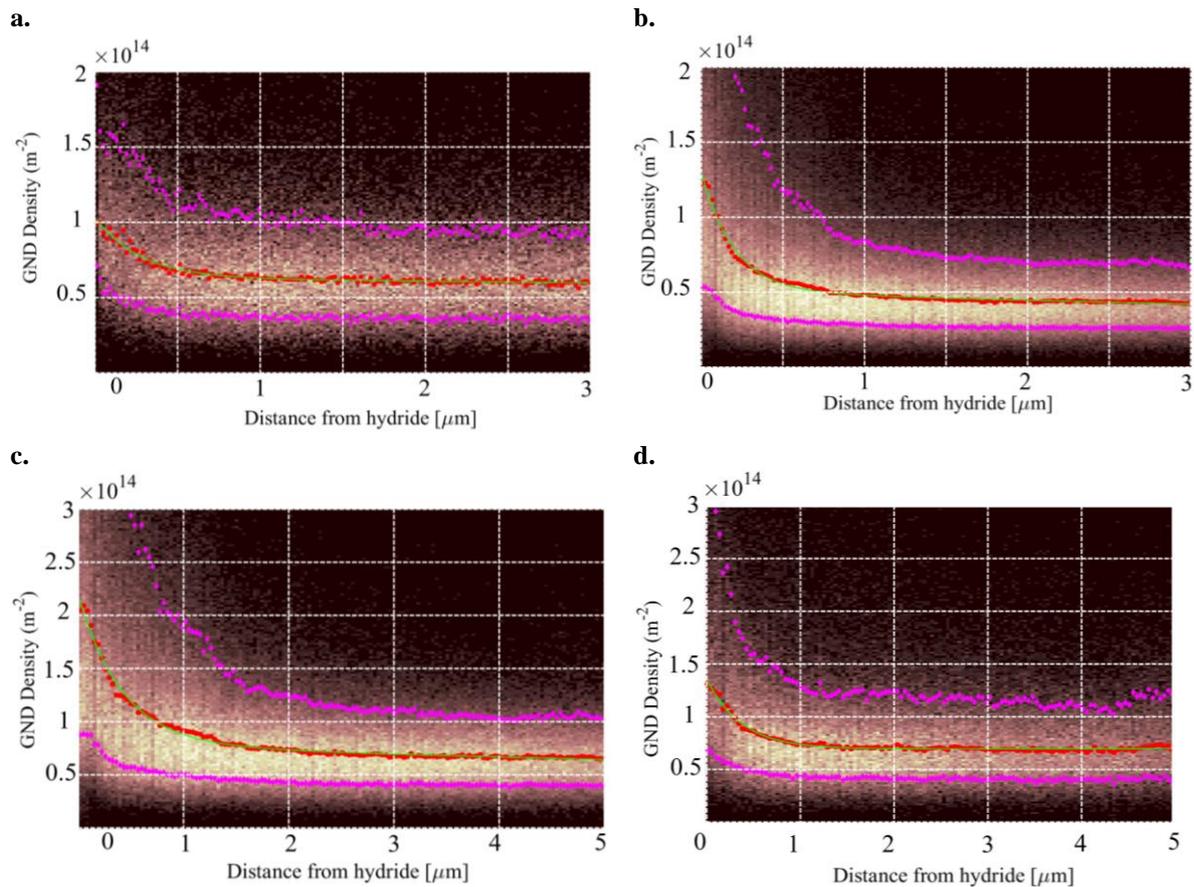

**Fig. 17. The spatial variations of GND densities with distancing from hydride-zirconium interfaces: (a) Two interacting hydrides, (b) interacting packs of hydrides, (c) grain boundary hydrides, and (d) triple point hydrides. GNDs are measured within the neighbouring α-grains. The grayscale color gradient indicates the frequency of mreasured GND densities for each distance. Red points represent medians while magenta points represent 10% and 90% values at each distance. Green lines are the fitted lines with the paramater values provided in Table-2.**

There is a clear increase in average GND densities close to hydrides. While there is a reasonable consistency from map to map for average GND densities, it is evident that both magnitude and the rate of increase in GNDs are more pronounced close to thicker hydrides. In addition, the magnitude of GNDs is the highest for triple point as well as intergranular hydrides. While GND concentration settles in ~1µm from the hydride interfaces in intragranular (or interacting) hydrides, this value for grain boundary hydrides is ~2µm. The significance of this number is in its role for designing microscale nano-indentation experiments on hydride-zirconium pillars or cantilevers [24,39]. This finding suggests that when the micromechanics of hydrides is studied using microcantilever beams or micropillars, it might be necessary to select cross sections or the dimensions of specimens bigger than 2µm to avoid perturbing the distribution of dislocation densities. Although, as it is evident from the results of previous section, all stress components damp at much higher distances. Hence making such small samples results in the release of, at least a portion of, residual stresses induced by hydride transformation strain.

**Table 2-** Parameters representing the variations of GND densities as a function of distance $x$ from the hydride interface with the surrounding α-grains. The implemented function is $\rho_{GND} = Ae^{-Bx} + C$.

| $\rho_{GND} = Ae^{-Bx} + C$ | A | B | C | Hydride thickness (μm) |
|---|---|---|---|---|
| **Two interacting hydrides** | 4.1e13 | 3.7 | 6.3e13 | 0.4 |
| **Pack of interacting hydrides** | 6.4e13 | 2.8 | 4.1e13 | 1.9 |
| **Intergranular hydrides** | 1.4e14 | 2.5 | 7.7e13 | 2.9 |
| **Triple point hydrides** | 7.1e13 | 3.3 | 7.1e13 | 2.5 |

So far, the spatial variations of GND densities within α-grains are discussed, however, both modeling and experimental results presented in sections 3.3 and 3.4 appear to show higher GND densities within hydrides. To further investigate, the average GND densities measured in the three intergranular hydrides of section 3.3 and those of triple point hydride of section 3.4 are calculated and presented in Table 3. Two approaches are used to compare the average GNDs from hydrides to those from surrounding α-grains. In the first approach, for calculating the average GNDs in α-grains, all points located in <1μm from the hydride-zirconium interface are used, while in the second approach, all point with GNDs>$10^{14}$ m$^{-2}$ are considered. This is done to account for grain size effects and that α-grains are bigger with more areas of low GND densities. It is evident from Table 3 that, regardless of the method used for calculating average GNDs within α-grains, the measured GND density in hydrides are always higher. This is another important evidence that a portion of hydride transformation strain is accommodated by the plastic deformation inside hydrides [87,88].

**Table-3.** Average GND densities (m$^{-2}$) in hydrides and surrounding α-grains.

| Map | Intergranular ROI$_1$ | Intergranular ROI$_2$ | Intergranular ROI$_3$ | Tripple point hydride |
|---|---|---|---|---|
| **Hydride** | 4.13e14 | 3.02e14 | 4.81e14 | 2.15e14 |
| **α-zirconium (GND>$10^{14}$ m$^{-2}$)** | 2.24e14 | 2.30e14 | 2.15e14 | 1.44e14 |
| **α-zirconium (d<1μm)** | 2.10e14 | 2.12e14 | 1.95e14 | 1.08e14 |

Fig. 18 represents the distribution of total plastic shear strain accommodated by all slip systems of neighboring α-grains as a function of the distance from hydride-zirconium interfaces. These values are computed by CPFE. For the sake of clarity, only the regions with a total shear strain exceeding 0.001 are shown. Fig. 18a and 18b display the variations for two interacting hydrides and interacting packs of hydrides, respectively, while Fig. 18c and 18d display such variations for intergranular and triple point hydrides, respectively. Included in the figures are the 10%, 50%, and 90% of the calculated values, while respecting the size of the contributing element. In addition, the calculated relative activity of each slip system is provided in Fig. 18e. Relative activities are calculated by dividing the total shear on each slip set to the total cumulative shear on all slip systems. In all shear calculations, the absolute values are used to avoid cancellation of negative shears by the positive ones.

Generally, shear strains are notably higher in the vicinity of thicker hydrides. Although shear strains typically diminish with distancing from the hydride interfaces, for interacting hydrides (Fig. 18a) an increase is observed in the midline, underscoring the pronounced effect of hydride-hydride interactions. Further, it is evident from Fig. 18e that the basal slip system is the most active system in all cases, even though slip on prism has the lowest CRSS. This is consistent with the results of previous studies, conducted by molecular dynamics modeling, that show the accumulation of dislocations on the basal slip systems [51]. The second most active slip system is prism followed by pyramidal for some cases, yet both activities are a lot smaller than that of basal.

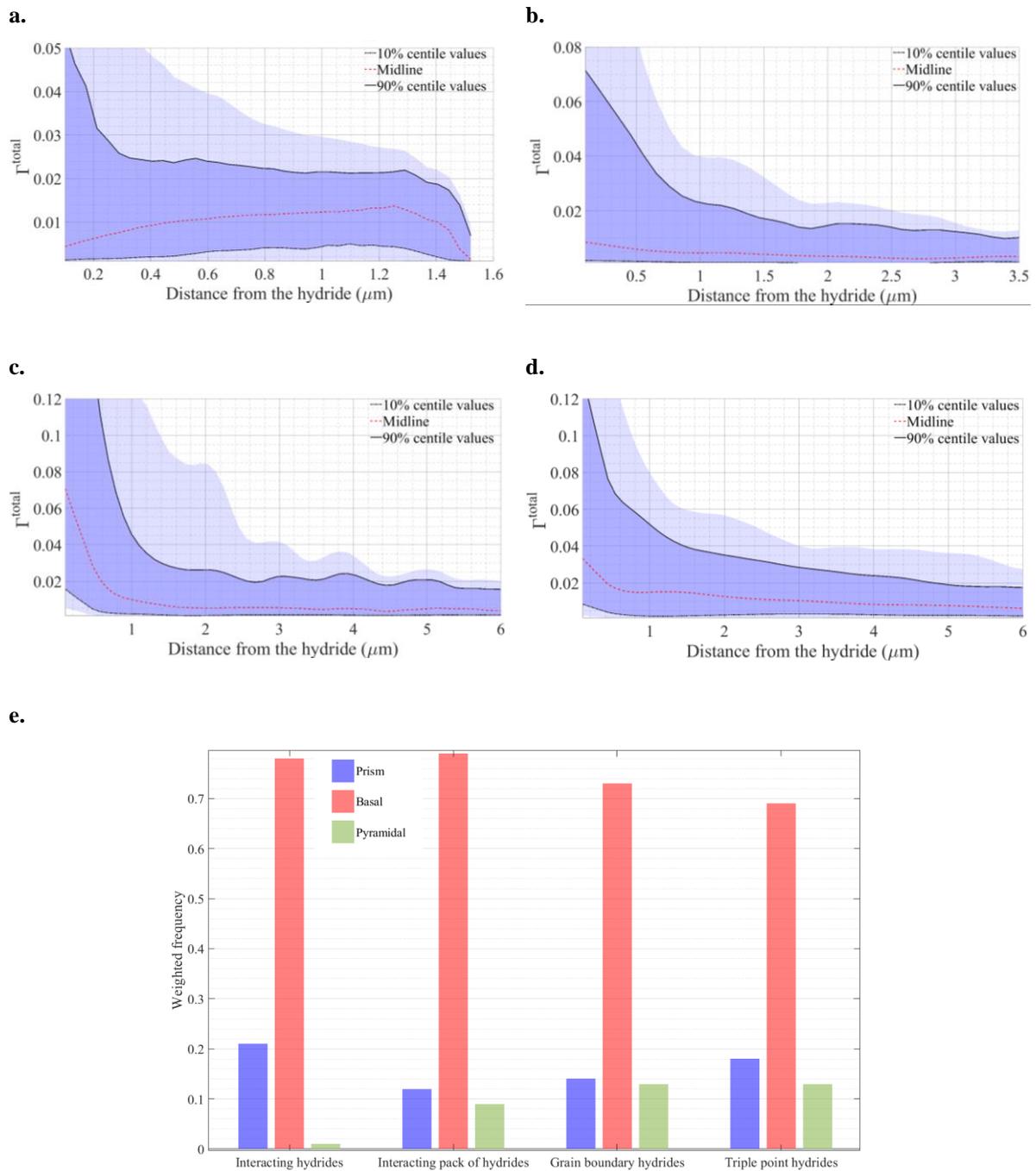

**Fig. 18. The total cumulative plastic shear accommodated on all slip systems within the neighboring α-grains as a function of distance from hydride-zirconium interfaces: (a) Two interacting hydrides, (b) interacting packs of hydrides, (c) grain boundary hydrides, and (d) triple point hydrides. The distributions include 10%, 50%, and 90% of the calculated values. (e) The relative activity of each slip system.**

### 4.4. The cumulative effects of hydride interactions

The role of hydride interactions in the development of localized deformation zones is further studied in this section. The two interacting hydrides of section 3.1 are analyzed using two CPFE simulations. As shown in Fig. 19a, in the first simulation, the left-hand side hydride (hydride B) is allowed to precipitate in the absence of the right-hand side hydride (hydride A), i.e., the zone representing hydride A is replaced by the parent α-grain. In the second simulation, hydride B was allowed to precipitate in the absence of hydride A. The rotations calculated from the first and second simulations are respectively shown in Fig. 19a and 19b, while those calculated for the simultaneous precipitation of both hydrides are shown in Fig. 19c. It is evident from the results that a relatively symmetric rotation field develops around the tip of the hydride B. This symmetric rotation field is highlighted by the red and yellow ovals of Fig. 19a. Similarly, the results of the second simulation reveal a symmetric rotation field around hydride A. However, when the two hydrides are precipitated, the rotation field in between the two hydrides is positively resonated, while the one outside of the zone is annihilated. That is, the rotation field calculated in the red oval of Fig. 19a is annihilated in Fig. 19c, while the one calculated in the yellow oval is significantly increased. In fact, the precipitation of the two parallel hydrides has resulted in a polarized rotation field in favor of the area confined in between the two.

Consistent with the results of the previous sections, the most active slip system is basal in all simulations of this section. The schematic of the plane normal and shear direction of this slip system is shown in Fig. 19a. The resolved shear stresses (RSS) calculated on the most active basal variant are shown in Fig. 19d-19f. The sequence of figures follows the one described for the rotation fields. Similar to the calculated rotation fields, it is evident that in the region confined in between the hydrides, resolved shear stresses from hydrides are directionally aligned, however, moving beyond this zone, RSS form the two hydrides counteract due to having opposite signs leading to the annihilation of RSS. These areas are highlighted by the two red and yellow ovals of Fig. 19f.

Fig. 19g and 19h further illustrate the distribution of resolved shear stress and calculated rotations along path #2 shown in Fig. 19d. The rotation field calculated for the region confined in between the two hydrides is significantly higher than the summation of the rotation field induced by the precipitation of each hydride. In contrast, outside the confined zone, the combined rotation field is less than the one induced by each hydride due to the opposing RSS fields. It is worth mentioning that the precipitation of both hydrides leads to a significantly higher localized stresses which subsequently induce higher RSS acting on the basal slip systems and higher rotation fields.

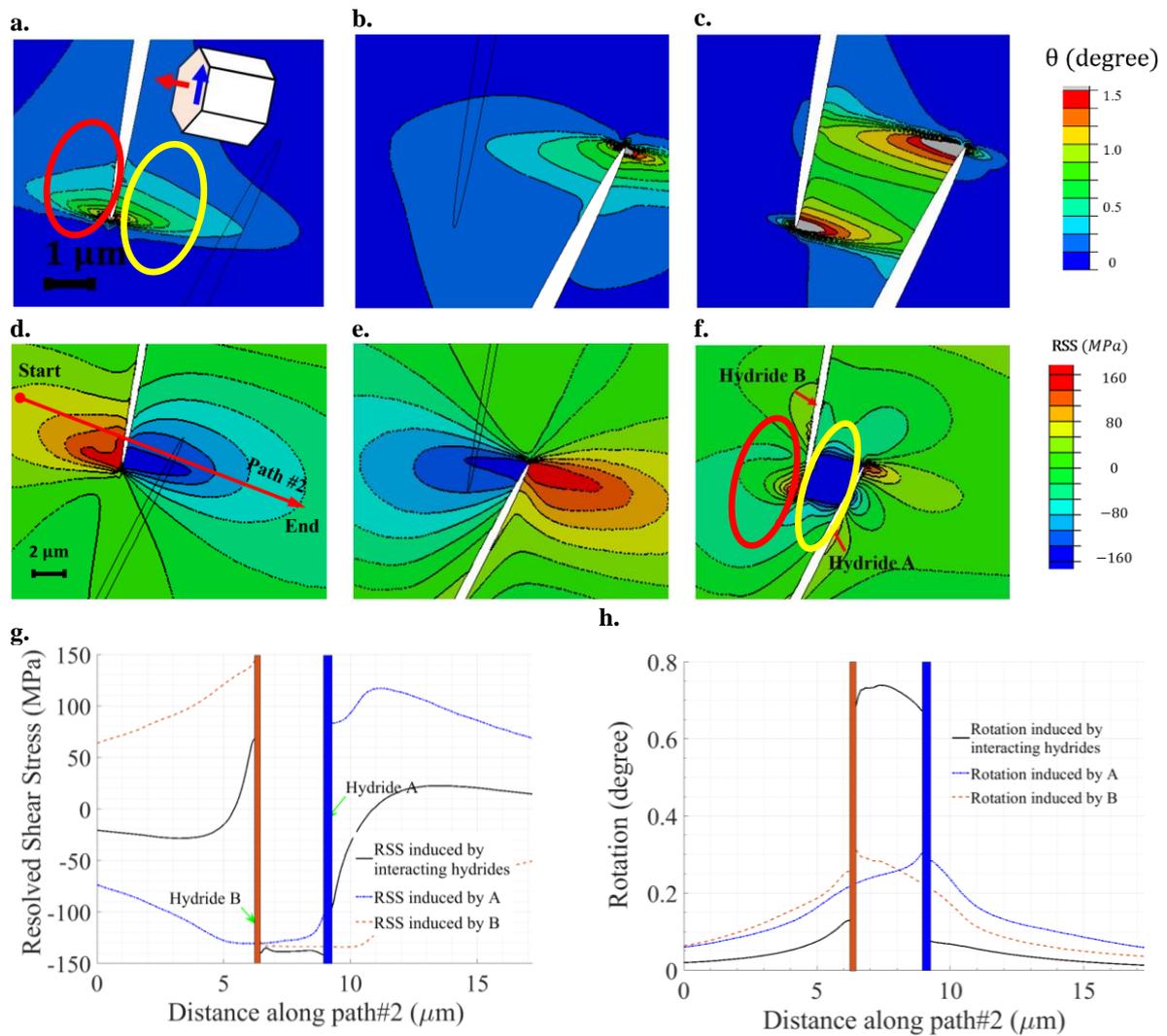

**Fig. 19.** The cumulative effects of hydride interactions: (a) The rotation field induced by the precipitation of hydride B in the absence of hydride A, (b) the one induced by the precipitation of hydride A in the absence of hydride B, and (c) the rotation field induced by the precipitation of both hydrides. The corresponding resolved shear stresses (RSS) acting on the most active basal slip system are respectively shown in (d) to (f). (g) and (h) The variation of RSS and rotation along the path #2 shown in d. Hydride A and B are labeled in (f).

### 4.5. Memory effects and thermal cycling

Understanding the evolution of stress and dislocation fields during precipitation, dissolution, and reprecipitation of hydrides undergoing thermal cycles is very important as such cases happen for the zirconium-alloy components used in the core of nuclear reactors. During shut down for service or maintenance, the temperature of reactor core components decreases leading to the precipitation of hydrides if solid solution solubility limit of hydrogen in zirconium is exceeded. Conversely, as the temperature of core components increases while going to full power, a fraction, or all, of hydrides may dissolve. Over the lifetime of nuclear reactors, core components will undergo several thermal cycles. Hence, in this section, we focus on studying this process. So far, we showed that the method implemented in the CPFE code for simulating hydride precipitation provides reasonable results for rotation and GND fields when compared with the measured ones. Here, attention is given to CPFE predictions for both GND and rotation fields as well as hydrostatic stress fields under thermal cycling.

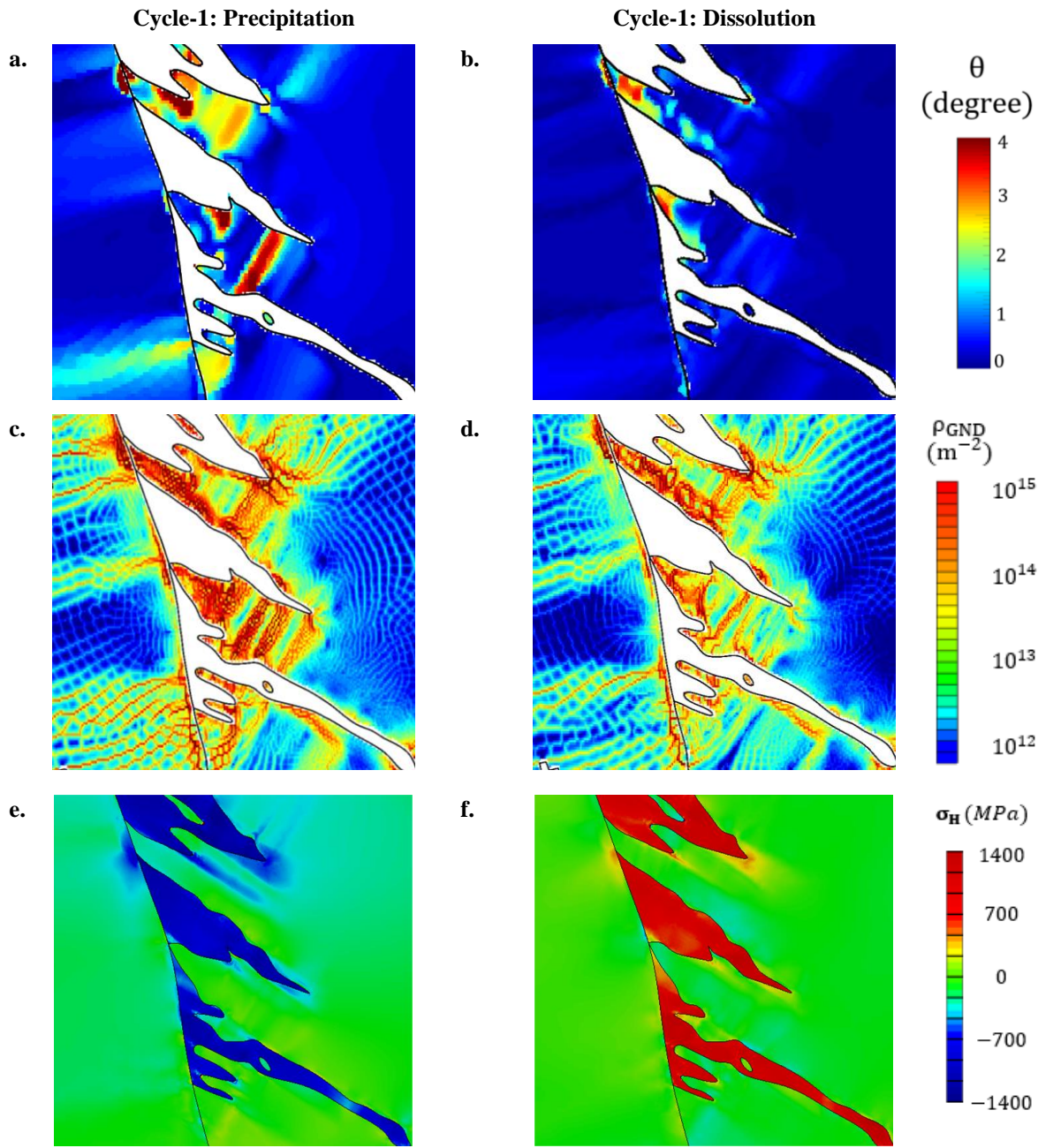

**Fig. 20. Memory effects during hydride precipitation and dissolution. CPFE results in the left column are for precipitation and those presented in the right column are for subsequent dissolution. (a, b) Rotation fields, (c, d) GND densities, and (e, f) hydrostatic stresses. All results are from CPFE simulations and are for the ROI$_3$ of grain boundary hydrides of section 3.3.**

First, precipitation and dissolution in one thermal cycle is studied. Specifically, we focus on the case of intergranular hydrides of section 3.3, ROI$_3$. Hydride precipitation was modeled using the same methodology described in section 2, while for dissolution, the sign of hydride transformation strain was reversed and was subsequently applied to the same hydride domain. By doing so, the transformation strain that was applied for precipitation was effectively removed during dissolution. Fig. 20a and 20b respectively show the calculated rotation field within zirconium α-grains after the precipitation and dissolution of hydrides. The corresponding GND and hydrostatic stresses are respectively shown in the second and third rows of Fig. 20.

It is shown in Figs. 20a and 20b that even after full dissolution of hydrides, there exists a residual rotation field which necessitates the presence of residual GNDs shown in Fig. 20d. While the magnitude of both rotation and GND densities are lower after dissolution, the presence of GNDs help with re-nucleation of hydrides from the same location. Generally, dislocations trap hydrogen atoms leading to local increase in hydrogen concentration and the subsequent hydride nucleation. This phenomenon is known as memory effects [89,90]. Most importantly, while the calculated hydrostatic stress within the hydride is compressive (Fig. 20e) after precipitation, a very tensile hydrostatic stress field is calculated for the same domain after dissolution. Tensile hydrostatic stresses lead to further accumulation of hydrogen atoms and the re-precipitation of hydrides in the same domain. CPFE results presented here are in very good agreement with the ones obtained by in situ TEM experiments recently reported by Hanlon et al [90] for precipitation, dissolution and reprecipitation of hydrides in zirconium. In addition, with the use of discrete dislocation plasticity, Patel et al. [57] calculated a tensile hydrostatic stress field in the dissolved hydride domain. It was suggested that the presence of dislocation network near the hydride nucleation site contributes significantly to the generation of this tensile stress. While our CPFE method does not directly incorporate the effects of individual dislocations on the development of tensile hydrostatic stresses, we show that the constraints imposed by the surrounding α-grains on the dissolved hydride domain can also be a major contributing factor on the development of such stress field further contributing to the memory effects.

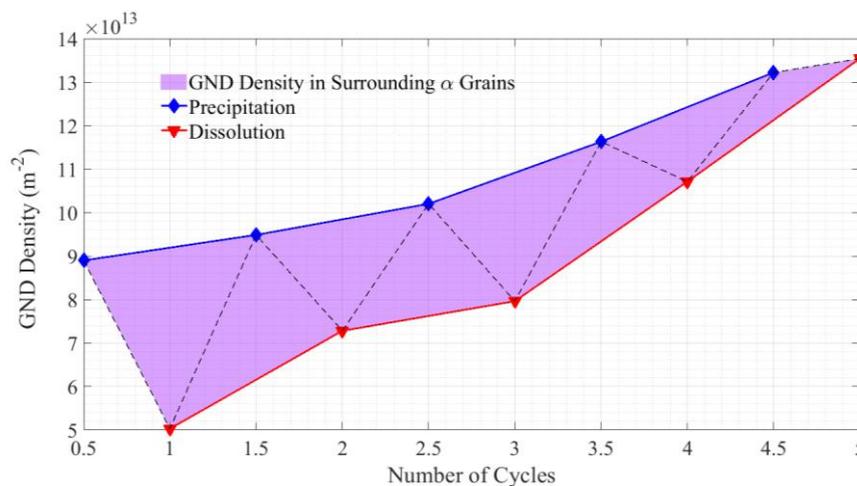

**Fig. 21. The effects of thermal cycling on precipitation, dissolution, and re-precipitation of hydrides: The evolution of GND densities in zirconium α-grains. Results are from CPFE simulations and are for ROI$_3$ of grain boundary hydrides of section 3.3.**

In addition to dissolution, the effects of thermal cycling are also investigated in this section. Such effects were replicated by applying positive and negative hydride transformation strain to the hydride domain so that precipitation, dissolution, and re-precipitation can be further studied. Fig. 21illustrates the evolution of average GND density in the surrounding zirconium α-grains. A clear increase in the dislocation density is observed with cycling. Further, while after each dissolution GND density in the matrix decreases, the difference between GND densities after precipitation from those after dissolution decreases. That is, the effects of thermal cycling are noticeable in the first few cycles and diminish with further cycling. In addition, as GND density increases, the surroundings α-grains will work harden leading to the development of higher localized stresses with further cycling. The observed trends of this section are in very good agreement with the ones obtained using a dislocation dynamic model [53,57], where it was shown that the rate of increase in the dislocation density decreases with further thermal cycling.

## 5. Conclusions

The rotation and GND fields induced by the precipitation of zirconium hydrides in the absence of any external mechanical load were measured by EBSD and simulated by CPFE modeling. Attention was given to interacting intragranular hydrides, packs of hydrides, grain boundary hydrides and triple point hydrides. All hydrides were shown to be FCC δ-hydride that maintain a crystallographic orientation relationship with the surrounding HCP α-grains. It is shown that:

(1) The precipitation of hydrides induces large lattice rotations and highly localized GND fields both in the surrounding α grains and the hydride itself. The magnitudes of lattice rotations and GNDs can exceed $4°$ and $10^{15}$ $m^{-2}$, respectively.
(2) The magnitudes of both rotation and GND fields are much higher in the vicinity of hydride tips, both in the surrounding α grains and the hydride itself.
(3) In comparison to the surrounding α-grains and for the hydrides studied here, on average, higher GND densities are measured and calculated within hydrides.
(4) The strength of GND density fields diminishes rapidly by distancing from the hydride-zirconium interfaces. While GND concentrations settles in ~1µm from the hydride interfaces in intragranular (or interacting) hydrides, this value for grain boundary hydrides is ~2µm. For the cases analyzed here, it was shown that this settling distance depends on the thickness of hydrides.
(5) The effects of α-grain orientations as well as those of hydrides were shown to be significant where polarized rotation, GND, and stress fields were observed to develop for interacting hydrides as well as hydride packs. The calculated stress fields are shown to explain the configuration of hydrides and why they grow axially.
(6) The polarized rotation field measured and calculated in the confined area in between interacting hydrides is higher than the summation of the rotation fields induced by individual hydrides if each was precipitated in the absence of the other. Numerical results revealed that this is due to the patterning of resolved shear stresses acting on the slip systems of surrounding α-grains.
(7) Both CPFE and EBSD results revealed the development of discrete and parallel GNDs, which are at their maximum in the vicinity of hydrides. These fields are shown to form due to the slip activity that primarily takes place on basal slip system.
(8) Significant stress and rotation variations are observed within α-grains in between hydrides, as well as across hydrides and surrounding α-grains.
(9) CPFE simulations revealed that hydride domains respectively develop compressive and tensile hydrostatic stresses during precipitation and the subsequent dissolution cycle. The magnitude of localized GND densities in the vicinity of hydride-zirconium interface increases with thermal cycling, which in combination with the patterning of hydrostatic stress, leads to the re-precipitation of hydrides from the same location. CPFE results revealed that thermal cycling effects are noticeable in the first few cycles and weaken with further cycling.

**Acknowledgment**

This work was supported by an Alliance Grant (ALLRP 560391 - 20) as well as a Discovery Grant (RGPIN- 2022-02955) from the Natural Sciences and Engineering Research Council of Canada (NSERC). In addition, this research was undertaken, in part, thanks to funding from the Canada Research Chairs Program.

**Contributions**

**Masoud Taherijam :** Conceptualization, Formal analysis, Investigation, Methodology, Visualization, Writing – original draft. **Saiedeh Marashi :** Methodology, Formal analysis, Investigation. **Alireza Tondro :** Investigation. **Hamidreza Abdolvand :** Supervision, Conceptualization, Formal analysis, Funding acquisition, Resources, Writing – review & editing.

**Data availability**

All EBSD maps and measured Euler angle files (CTF EBSD files) in support of this paper are available at Zenodo.org.